\documentclass[pre,twocolumn,a4paper]{revtex4-1}
\usepackage{graphics}
\usepackage{amssymb}
\usepackage{amsmath}
\usepackage{wasysym}
\usepackage{latexsym}
\usepackage{graphicx}
\usepackage{epsfig}
\usepackage[caption=false]{subfig}
\begin{document}

\title{The spatial dynamics of ecosystem engineers }

\author{Caroline Franco and   Jos\'e F. Fontanari }     
\affiliation{Instituto de F\'{\i}sica de S\~ao Carlos,
  Universidade de S\~ao Paulo,
  Caixa Postal 369, 13560-970 S\~ao Carlos, S\~ao Paulo, Brazil}

\begin{abstract}
The changes on  abiotic features of ecosystems  have rarely been taken into account by population dynamics models, which 
 typically focus  on trophic and competitive interactions between species. However, understanding the population dynamics of organisms that must modify their habitats in order to survive, the so-called ecosystem engineers, requires the explicit incorporation of abiotic interactions in the models.
Here  we study a model  of ecosystem  engineers that is discrete both in space and time, and where the engineers and their habitats are arranged in patches fixed to the sites of regular lattices. The growth of the engineer population is modeled by Ricker equation with a density-dependent carrying capacity that is given by the number of modified habitats. A diffusive dispersal stage ensures that a  fraction of the engineers move from their birth patches to neighboring patches. We find that dispersal influences the metapopulation dynamics only in the case that the local or single-patch dynamics exhibit chaotic behavior. In that case, it can suppress the chaotic behavior  and avoid  extinctions in the regime of large intrinsic growth rate of the
population.
\end{abstract}

\maketitle

\section{Introduction}\label{sec:intro}

The usefulness of the  concept of ecosystem engineers as organisms that change their environment by causing physical state changes in living or non-living materials  \cite{Jones_94,Gurney_96} is  somewhat controversial since all species modify their environment \cite{Reichman_02}.  However,  species that must modify their habitats in order to survive or that  increase significantly their chances of reproduction and survival  in the modified habitats certainly comprise a unique class of organisms. These benefits, in addition, are likely to alter their population dynamics, which may require the explicit incorporation of abiotic interactions \cite{Cuddington_09}.  For instance, beavers (\textit{Castor canadensis}) are considered  as model  systems of ecosystem engineers \cite{Wright_04}, since  the  areas flooded by the beaver dams  increase the distance they  can travel by water, which is safer than traveling by land, and results in a net increase of their survival expectations   \cite{Dawkins_82}. 

A key feature of the population dynamics of ecosystem engineers is that the growth of the  engineer population is limited by the number of usable habitats, which in turn are created by the engineers via the conversion of virgin habitats. This feedback loop results in a density-dependent carrying capacity that is behind the peculiar dynamical behavior exhibited by the
mathematical models of those systems. To our knowledge,  the first  mathematical model of the population dynamics of ecosystem engineers that  explicitly takes into account the interactions between the organisms and the habitats  was put forward by Gurney and Lawton in the mid 1990s \cite{Gurney_96}.  In that model, the quality of the habitats takes on three different discrete states: virgin, usable (or modified) and degraded habitats. The  role of the  engineers is to effect the transition from  the virgin to the usable  states.  The modified state then  transitions to the degraded state, which is unsuitable for occupation by the engineers. Finally, the degraded state recovers to the virgin state that can then be reused  by the engineers. Analysis of the  continuous-time population dynamics model shows the existence of fixed points characterized by the presence of all three habitat states as well as cycles  in which the number of engineers and the number  of virgin habitats 
oscillate out of phase  \cite{Gurney_96}. 

 In this contribution we consider a spatial version of Gurney and Lawton model in which the engineers and their habitats are arranged in patches fixed to the sites of regular lattices. A diffusive dispersal stage ensures that a  fraction $\mu$ of the engineers move from their birth patches to neighboring patches \cite{Hassell_91,Comins_92}.  By patch  we  mean  an  ecosystem  composed of all three types of habitats as well as of the  engineer population. Hence our approach differs starkly from a previously studied patch dynamics model in which 
 each patch represents an habitat state and  only the transitions between the habitat states  are considered, i.e., the dynamics of the engineer population are not accounted for \cite{Wright_04}. To facilitate the analysis of the discrete patch  model we make time discrete as well and replace the logistic growth equation of the original model by the Ricker equation \cite{Ricker_54,Murray_03a} so that the spatial 
model reduces to  a coupled map lattice \cite{Kaneko_92}. Henceforth we will refer to the system of patches as metapopulation.

We find that the diffusive dispersal stage influences the metapopulation dynamics only in the case the local or single-patch dynamics exhibits chaotic behavior. In that case, we find that for certain values of the dispersal fraction $\mu$ the chaotic behavior  is suppressed and the dynamics enter a two-point cycle in which the engineer densities  at each patch oscillate between high and low values, forming rich (two-dimensional) geometric patterns. Another interesting effect of the dispersal is the avoidance of extinction, which is one of the  outcomes of the local dynamics when  the  intrinsic growth rate $r$ of the engineers is large.  In fact, the  unexpected possibility of extinction of the population for large $r$ is a consequence of the density-dependent  carrying capacity and the study of the spatial dynamics of ecosystem engineers offers a neat example of  how multiple populations coupled by diffusive dispersal can  eliminate or reduce that risk.

The rest of the paper is organized as follows. In Sect.\ \ref{sec:model} we introduce our time discrete version of Gurney and Lawton model of
ecosystem engineers and derive the recursion equations of the local or single-patch dynamics for the  density of engineers as well as for the fractions of the three types of habitats. In that section we introduce also  the equation that describes the  diffusive dispersal stage  of the engineer population. The local dynamics is then studied in great detail in Sect.\ \ref{sec:local} with emphasis on the stability analysis of the fixed points.  The spatial model with the patches  arranged in a chain and a square lattice  is considered in Sect.\  \ref{sec:spatial}, where the
dynamic behavior of the metapopulation is studied  mainly through the analysis of bifurcation diagrams. Finally, Sect.\ \ref{sec:conc} is reserved to our concluding remarks.

\section{Model}\label{sec:model}

Here we build on the model of Gurney and Lawton \cite{Gurney_96} to study the dynamics of a population of organisms that must modify their own habitat in order to survive.  Let us first rewrite  the original, continuous-time model  as a discrete-time model. We assume that
the population of engineers at generation $t$ is composed of $E_t$ individuals and that each individual requires a unit of usable habitat to survive. In addition, we denote the  number of units of usable habitats available at generation $t$  by $H_t$. Since $H_t$ plays the role of a time-dependent carrying capacity for the population of engineers, we can use Ricker model   to write the expected number of engineers at generation $t+1$ as  
\begin{equation}\label{EE1}
E_{t+1} = E_t \exp \left [ r \left ( 1 - E_t/H_t \right ) \right ],
\end{equation}
where  $r$ is the intrinsic growth rate  of  the population of engineers \cite{Ricker_54,Murray_03a}. 

What makes the model  interesting is the requirement that usable habitats be created by engineers working on virgin habitats. More pointedly, if  $V_t$ denotes the units of virgin habitats at generation $t$, then a portion $C \left ( E_t \right ) V_t$ of them will be  transformed in usable habitats at the next generation, $t+1$. Here  $C \left ( E_t \right )$ is any function that satisfies $ 0 \leq C \left ( E_t \right )  \leq 1$ for all $E_t$ and $C \left ( 0 \right ) = 0$. However, usable habitats do not last forever and eventually  decay to degraded habitats which are useless to the engineers. Denoting by $\delta H_t$ the portion of usable habitats that
decay to degraded habitats in one generation,  we can immediately  write an equation for the expected number of units of usable habitats at generation $t+1$,
\begin{equation}\label{HH1}
H_{t+1} = \left ( 1 - \delta \right ) H_t + C \left ( E_t \right ) V_t,
\end{equation}
where $ \delta \in \left [ 0, 1 \right ]$ is the decay fraction. Degraded habitats will eventually recover and become virgin habitats again.  If we denote by $\rho D_t$ the portion of degraded habitats that recover to virgin habitats we have 
\begin{equation}\label{DD1}
D_{t+1} = \left ( 1 - \rho \right ) D_t + \delta H_t,
\end{equation}
where $ \rho \in \left [ 0, 1 \right ]$ is the recovery fraction. Finally, the recursion equation for the expected number of units of virgin habitats is simply
\begin{equation}\label{VV1}
V_{t+1} =    \left [ 1 - C \left ( E_t \right ) \right ]  V_t +  \rho  D_t .
\end{equation}
We note that $ V_{t+1} + H_{t+1}  + D_{t+1} = V_{t} + H_{t}  + D_{t}   = T$ so that the total supply of habitats $T$ is fixed. This remark motivates the introduction of the habitat fractions $v_t \equiv V_t/T$, $h_t \equiv H_t/T$ and $d_t \equiv D_t/T$ that satisfy $v_t + h_t + d_t =1$ for
all $t$. In addition, we introduce the density of engineers $e_t = E_t/T$ which, differently from the habitat fractions, may take on values greater than 1.
In terms of these quantities, the recursion equations are rewritten as
\begin{eqnarray}
e_{t+1} &  =  & e_t \exp \left [ r \left ( 1 - e_t/h_t \right ) \right ]  \label{e} \\
h_{t+1} & = & \left ( 1 - \delta \right ) h_t + c \left ( e_t \right ) v_t  \label{h} \\
v_{t+1} & = &  \rho \left ( 1 - v_t -h_t  \right ) + \left [ 1- c \left ( e_t \right ) \right ] v_t, \label{v}
\end{eqnarray}
where we have used $ d_t = 1 - v_t - h_t$ and $c \left ( e_t \right ) \equiv C \left ( T e_t  \right )$. Here we will consider the function
\begin{equation}\label{c}
 c \left ( e_t \right ) = 1 - \exp \left ( - \alpha e_t \right ),
\end{equation}
where $\alpha > 0$ is the productivity   parameter, which measures the efficiency of conversion of virgin to usable habitats by the engineers. This function indicates that the engineers do not work independently on the construction of usable habitats, otherwise one should have 
$ c \left ( e_t \right ) \propto e_t$. This is actually the situation for small population sizes, where  $c \left ( e_t \right ) \approx  \alpha e_t$, but as the  population of engineers increases they begin to interact antagonistically (i.e. the production of  two engineers is actually less than the sum of their productions taken independently of each other) since  $c \left ( e_t \right ) /\alpha e_t < 1$ for all densities $e_t$.  

In fact, the abstract function $c \left ( e_t \right )$ incorporates, perhaps, the most interesting features of the system of engineers   as, for instance, the collaboration strategies  and the communication networks that allowed this class of organisms to build complex  collective structures (e.g., termite mounds and  anthills), which  may be  viewed as the organisms'  solutions to the problems that endanger their existence \cite{Fontanari_14,Francisco_16,Reia_16}. For humans,  this function  must incorporate the beneficial effects of the technological advancements  \cite{Petroski_94,Arthur_09} and this opens the interesting possibility of studying the interplay between technology evolution and population dynamics, similarly to what  has been done for genetics and culture  \cite{Boyd_05}.
It is also interesting to note that if one chooses the function   $c \left ( e_t \right )$  such that it vanishes  with the square of the density of engineers  for small population sizes, then one would observe an Allee effect, in which the population quickly goes extinct below a  critical population size or density \cite{Courchamp_08}. This choice  would correspond to the situation where the engineers are  obligate cooperators.

Although the decay of usable to degraded  habitats is  probably density dependent, particularly if the degradation results from the overexploitation of resources, here we  have modelled it by a constant probability $\delta$. There are examples, however, that conform to this assumption such as the case of  bark beetles that cannot breed in old dead wood so that  the usable habitats, namely, recently killed trees, degrade at a rate that does not depend on the presence of those insects \cite{Raffa_87}.

We refer to the system of  recursion equations (\ref{e})-(\ref{v}) as the local or single-patch population dynamics. By patch  we  mean  a  ecosystem, say an isle, with the three types of habitats and, possibly, a thriving population of engineers. Let us consider now a system of $N$ patches, say an archipelago, and assume that the engineers can diffuse to neighboring patches. More pointedly, we assume that a
fraction $\mu$ of the population of engineers  of  patch $i$ moves to the neighboring patches, so that after the dispersal stage the population at patch $i$ is  
\begin{equation}
E'_{i,t} = \left ( 1 - \mu \right ) E_{i,t} + \mu  \sum_j E_{j,t}/K_i,
\end{equation}
where  the sum is  over the $K_i$ nearest neighbors of patch $i$. Assuming that the  total supply of habitats $T$  is the same for all patches and dividing both sides of this equation by $T$ we get
\begin{equation}\label{el}
e'_{i,t} = \left ( 1 - \mu \right ) e_{i,t} + \mu  \sum_j e_{j,t}/K_i,
\end{equation}
which describes the effect of dispersal on the density of engineers in patch $i$. We note that in the case the patches have different sizes, which are gauged by the total supply of habitats $T_i$, we can still get an equation similar to eq.\ (\ref{el}) relating the engineer densities in  neighboring patches. The only complication is the appearance of the  ratios $T_j/T_i$ multiplying  the densities $e_j$ in the sum over the neighbors of patch $i$.
 The density after dispersal then follows the local dynamics  equations within each patch. We postpone to Section \ref{sec:spatial} the presentation of the complete set of equations, where  the habitat fractions exhibit the patch indices as well.

 In summary,  for each generation the dynamics consist of two phases: a dispersal phase and a  growing phase. In the dispersal phase, a fraction $\mu$ of engineers in each patch moves  to  neighboring patches according to eq.\ (\ref{el}), whereas in the growing phase
the engineers  reproduce and modify the habitat composition of their patches  according to the local dynamics equations  (\ref{e})-(\ref{v}).

\section{Local Population Dynamics}\label{sec:local}

In this section we study the system of  recursion equations (\ref{e})-(\ref{v}) and, as usual, we begin with the analysis of the 
 fixed-point solutions, obtained by setting $e_{t+1} = e_t = e^*$, $h_{t+1} = h_t = h^*$ and $v_{t+1} = v_t = v^*$. Then we consider the 
 oscillatory solutions, which are explored  in a somewhat more qualitative way through the analysis  of the  bifurcation diagrams. By oscillatory solutions we mean both periodic and chaotic solutions.

\subsection{Zero-engineers fixed point}

\begin{figure}[!ht]
\centering
\includegraphics[width=0.48\textwidth]{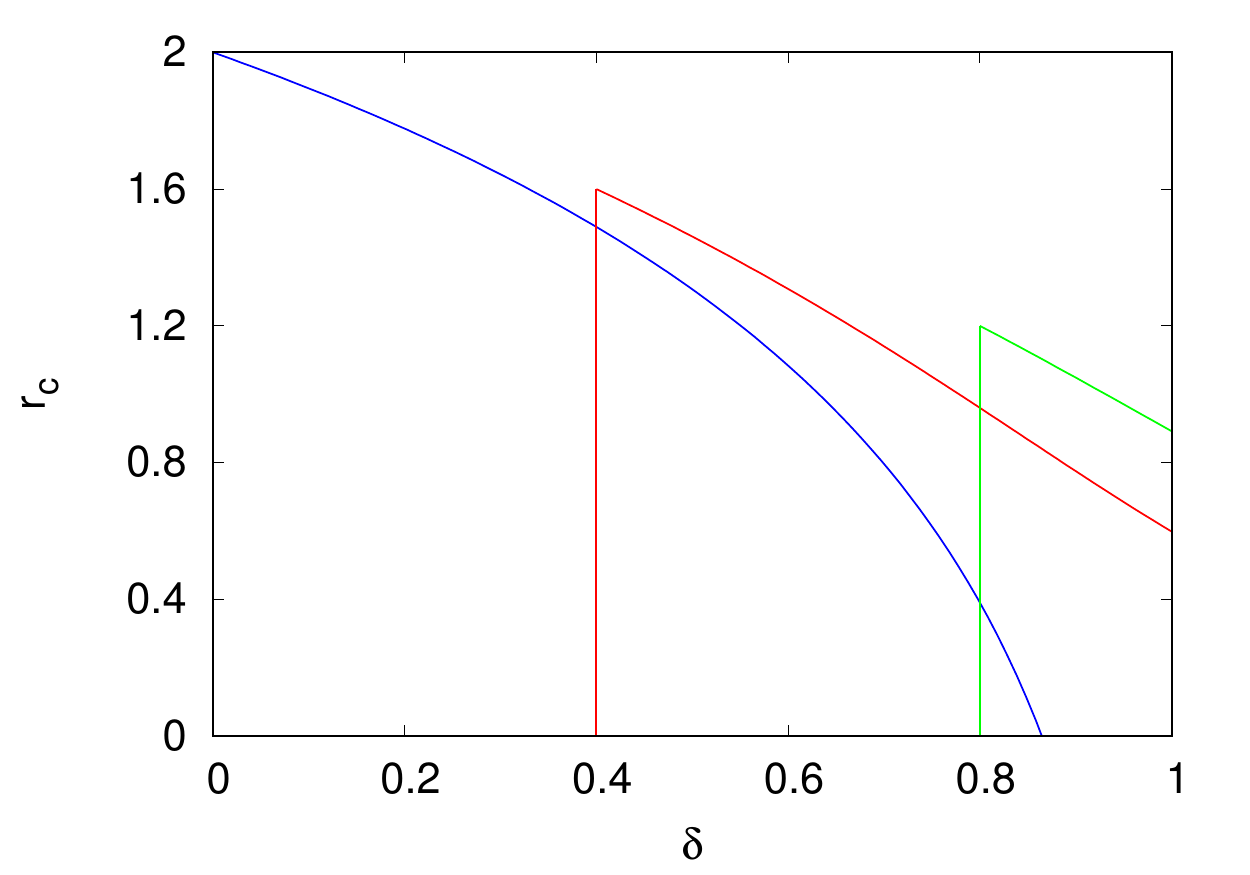}
\caption{(Color online) Intrinsic growth rate $r_c$ above which the zero-engineers fixed point is unstable  as function of the decay fraction  
$ \delta  $  for (top to bottom) $\alpha= 0.8, 0.4$ and  $0$.   For $ \delta < \alpha$ this fixed point is always unstable as indicated by the vertical lines in the figure.
For fixed $\alpha$,  the vertical line  intersects the $r_c \left ( \delta \right )$ line at the point  $\delta = \alpha  $  and $r_c = 2 - \alpha$. 
 }
\label{fig:rc_e0}
\end{figure}

Setting $ e^* = 0$ we obtain $h^* = 0$ and $v^* = 1$. This means that  in the absence of the  engineers, all usable habitats will degrade, then recover to virgin habitats, and stay forever  in that condition. The study of the local stability of this fixed point, however, is somewhat complicated because the ratio $x_t \equiv e_t/h_t$  is indeterminate. To circumvent this difficulty we rewrite the recursion equations (\ref{e})-(\ref{v}) in terms of the variables $x_t$, $e_t $ and $v_t$,
\begin{eqnarray}
e_{t+1} &  =  & e_t \exp \left [ r \left ( 1 -x_t \right ) \right ]  \label{e2} \\
x_{t+1} & = & \frac{x_t \exp \left [ r \left ( 1 -x_t \right ) \right ] }{  \left ( 1 - \delta \right )  +  x_t  v_t \hat{c} \left ( e_t \right )}  \label{x2} \\
v_{t+1} & = &  \rho \left ( 1 - v_t -e_t/x_t  \right ) + \left [ 1- c \left ( e_t \right ) \right ] v_t, \label{v2}
\end{eqnarray}
with the notation $\hat{c} \left ( e_t \right ) = c \left ( e_t \right )/e_t$, such that $ \hat{c} \left ( 0 \right ) = \alpha$.
At the fixed point  $x_{t+1} = x_t = x^*$ we have
\begin{equation}\label{x}
1 - \delta + \alpha x^* = \exp \left [ r \left ( 1 -x^* \right) \right ] .
\end{equation}
For $\delta = \alpha$ the solution is $x^* =1 $ and we can easily show that $x^* > 1$ for $ \delta > \alpha$ and $x^* < 1$ for $ \delta < \alpha$.   The local stability of the zero-engineers fixed point  is 
determined by requiring that the three eigenvalues of the Jacobian matrix
\begin{equation}\label{J0}
J_0 = 
\begin{bmatrix}
 \xi^*    &~~~~ 0   &~~~~ 0  \\
    (\alpha x^*)^2 /2 \xi^* &~~~~  1-x^* \left ( r + \alpha/\xi^*\right )  &~~~~ -\alpha (x^*)^2/\xi^* \\
    -\alpha - \rho/x^* &~~~~ 0 &~~~~ 1-\rho 
  \end{bmatrix}
 \end{equation} 
have absolute values less than 1. Here we  used the notation $\xi^* = \exp \left [r \left ( 1 -x^* \right) \right ]$.
The eigenvalues are $\lambda_1 = \xi^* $, $\lambda_2 = 1  - x^* \left ( r + \alpha/\xi^* \right ) $ and  $\lambda_3 = 1 - \rho$.
The condition $ \mid \lambda_1  \mid < 1$  is satisfied provided  that $x^* > 1$, which is guaranteed for  $\delta > \alpha$, whereas
the condition $ \mid \lambda_3  \mid < 1$ is always satisfied  since $\rho < 1$. However, the condition $ \mid \lambda_2  \mid < 1$ is violated
for  $r > r_c$,  where $r_c$ is determined by setting $\lambda_2 = -1$. Figure  \ref{fig:rc_e0} shows $r_c$ as function of $\delta > \alpha $ for fixed values of $\alpha$. In particular, for $\alpha=0$ we find $r_c = 2 + \ln \left ( 1 - \delta \right )$ so that for  $\delta > 1 - \exp \left ( -2 \right ) \approx 0.865$ the zero-engineers fixed point is unstable regardless of the value of $r$. We note that the recovery fraction $\rho $ has no effect whatsoever on the stability of the zero-engineers fixed point.

We stress that the violation of the condition $ \mid \lambda_2  \mid < 1$ results in the  oscillatory behavior of the ratio $x_t$ (hence the instability of the fixed point solution) and that the numerical iteration of the recursion equations  (\ref{e2}), (\ref{x2}) and (\ref{v2})  indicates that $e_t \to 0$ and $v_t \to 1$ in the asymptotic limit $t \to \infty$, despite the oscillatory behavior of $x_t$.  
Hence we conclude that  for $\delta > \alpha$ there is always a zero-engineers attractor (not necessarily a fixed point) characterized by $\lim_{t \to \infty} e_t = 0$, regardless  of the values of the parameters $\rho$ and $r$.  Somewhat surprisingly,  we will see in Section \ref{BD} that the  zero-engineers oscillating attractor plays a role in the dynamics  even for $\delta < \alpha$.

\subsection{Finite-engineers fixed point}\label{subsec:FE}

\begin{figure}[!ht]
\centering
\includegraphics[width=0.48\textwidth]{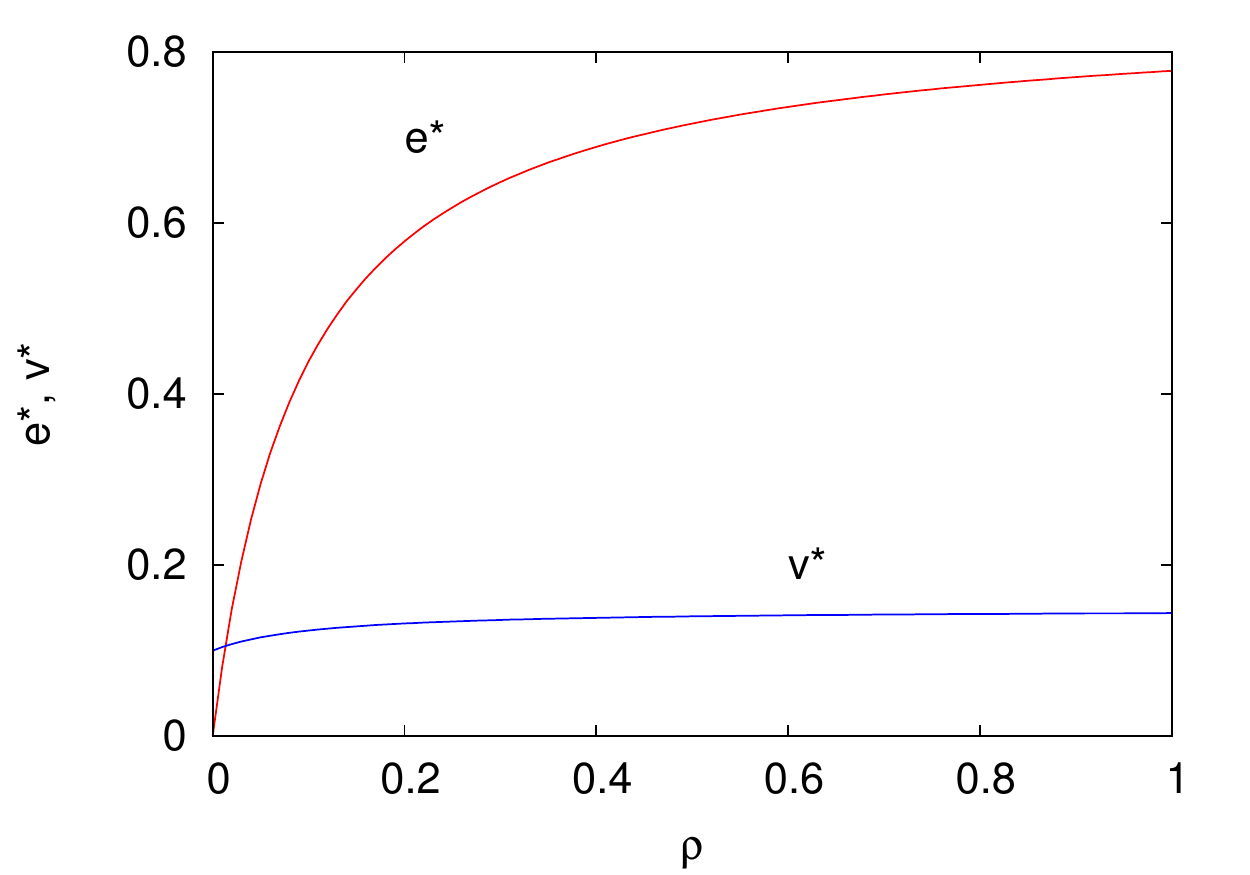}
\caption{(Color online) Finite-engineers fixed point $e^*$, $h^* = e^*$ and $v^*$ as function of the recovery fraction $\rho$ of degraded habitats for   
$\alpha = 1$ and $\delta= 0.1$. 
 }
\label{fig:0}
\end{figure}

Since $e^* > 0$ we have  $h^* = e^*$ (i.e., $x^* =1$) and $v^* = 1 - e^* \left ( 1 + \delta/\rho \right )$ with $e^*$ given by the solution of the  transcendental equation 
\begin{equation}\label{e*1}
\delta e^* = \left [ 1 - \exp \left ( - \alpha e^* \right ) \right ] \left [ 1 - e^* \left ( 1 + \delta/\rho  \right ) \right ] .
\end{equation}
For  $e^* \ll 1$  we find
\begin{equation}\label{e*2}
e^* \approx \frac{1 - \delta/\alpha}{1 + \delta/\rho}
\end{equation}
and $v^* = \delta/\alpha$, indicating that this fixed point is physical for $\delta < \alpha $ only. We find that the density of engineers at equilibrium increases monotonously from $e^* = 0$ at $\alpha = \delta $  to $e^* =  1/ \left ( 1 + \delta + \delta/\rho \right )$ as $\alpha \to \infty$. Moreover, as $\delta$ increases from $0$ to $\alpha$ , $e^*$ decreases monotonically from $1$ to $0$.  Finally, for small $\rho$ we
find $e^* \approx \rho \left ( 1/\delta - 1/\alpha \right)$ and $v^* \approx  \delta/\alpha$. As
$\rho$ increases, both $e^*$ and $v^*$  increase monotonically until some maximum values that depend on the parameters $\delta$ and $\alpha$, as illustrated in Fig.\ \ref{fig:0}. Interestingly,  although  $\rho$ is  the recovery fraction of degraded habitats that turn into virgin habitats, its increase  results only in a slight increment in the number of  virgin habitats. This is so because of the efficiency of the engineers in transforming virgin into usable habitats, which increases  to 1 exponentially fast with the  density of engineers, regardless of the value of the parameter $\alpha > \delta$.

The  analysis of the local stability of the finite-en\-gi\-neers fixed point  is facilitated if  we  use the variable   $x_t = e_t/h_t$ as before. 
The Jacobian matrix associated to the recursion equations (\ref{e2}), (\ref{x2}), (\ref{v2}) is
\begin{equation}\label{Je}
J_e = 
\begin{bmatrix}
 1    &~~~~ - r e^*   &~~~~ 0  \\
   -v^*  \hat{c}' \left ( e^* \right )  &~~~~  1-r -\delta &~~~~ - \hat{c} \left ( e^* \right ) \\
   -\rho  -\alpha  \left ( v^* - \delta  e^* \right )  &~~~~ \rho e^* &~~~~ 1 -\rho  - c \left ( e^* \right )  
  \end{bmatrix}
 \end{equation} 
with the notation $ \hat{c}'  ( e^* ) = d \hat{c}\left ( e \right )/d e \mid_{e=e^*}$, such that 
 $ \hat{c}'  \left ( 0\right )  = -\alpha^2/2$. This matrix reduces to $J_0$,  given by eq.\ (\ref{J0}), provided that  $e^* =0$ is the solution of the
 fixed point equation  (\ref{e*1}), viz. for $\alpha = \delta$, since  only in this case the zero-engineers fixed point has $x^* = 1$. 
 
 \captionsetup[subfigure]{labelformat=empty}
\begin{figure}[!ht]
\centering
\subfloat[$\rho = 0.1$] {\includegraphics[width=0.48\textwidth]{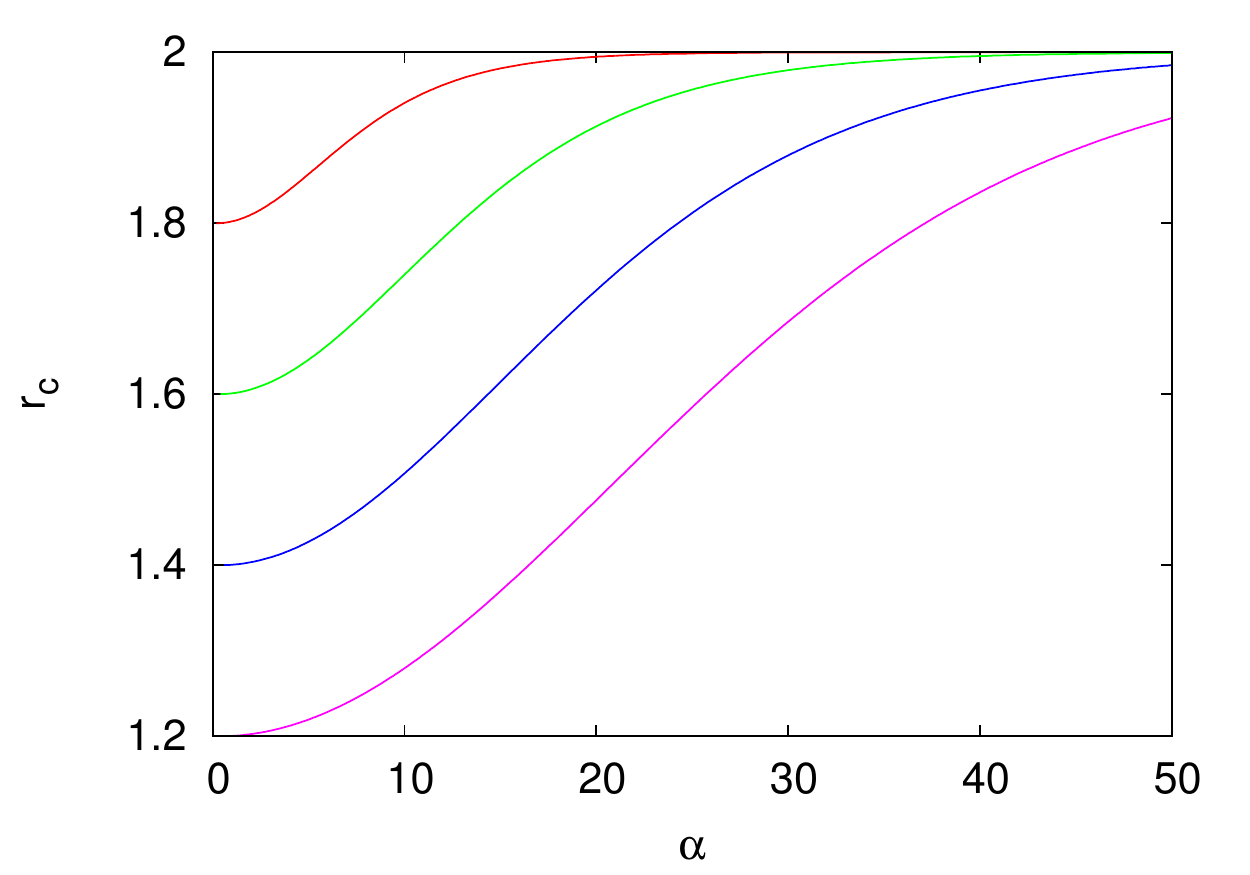}}\\
\subfloat[$\rho = 1$] {\includegraphics[width=0.48\textwidth]{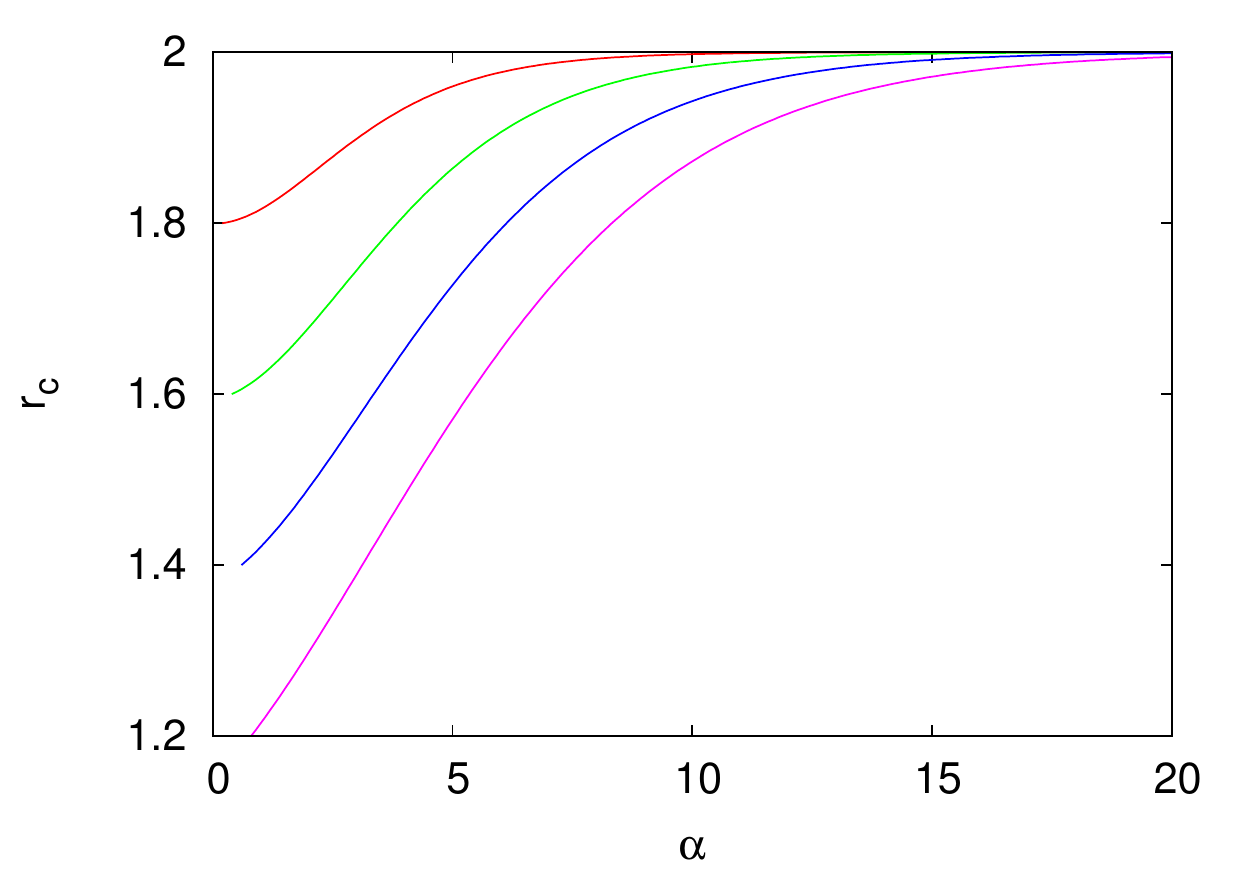}} 
\caption{(Color online) Intrinsic growth rate $r_c$ above which the engineer population exhibits oscillatory behavior as function of the parameter 
$\alpha > \delta $  for (top to bottom) $\delta= 0.2, 0.4, 0.6, 0.8$.  For $\alpha \approx \delta$ we have $r_c \approx 2 - \delta$ and for  $\alpha \to \infty$ we have $r_c \to 2$ regardless of the values of the parameters $\delta$ and $\rho$.
 }
\label{fig:1}
\end{figure}

 Analysis of the eigenvalues of the Jacobian matrix $J_e$  in the region $ \alpha > \delta$ where $e^* > 0$ (see eq.\ (\ref{e*2})) shows that 
 $\lambda_k < 1$ for $k=1,2,3$ (the labeling of the eigenvalues is the same as for the stability analysis of the zero-engineers fixed point).
However,  for large $r$  we find $\lambda_2 < -1$  thus  violating the condition of local stability of the finite-engineers fixed point. Since $e^*$ and $v^*$ do not depend on $r$ it is very easy to obtain numerically the value $r = r_{c}$ above  which the fixed point is unstable. Figure \ref{fig:1} shows the dependence of $r_c$ on the model parameters.  Regardless of the parameters we find that 
$r_c \in \left [ 2- \delta, 2 \right ]$. The lower bound $r_c = 2 - \delta$ is reached for $\alpha = \delta$ at which $e^* \to 0$. 
The upper bound $r_c = 2$ is reached in the limit $ \alpha \to \infty$. In fact,  since  in this limit
we have  $v^*/e^* = \delta$,  we can readily obtain $\lambda_2 =  1 - r$. Then setting $\lambda_2 =  -1$ yields $r_c = 2$. Actually,  
since $c \left ( e_t \right)  = 1$ in the limit $\alpha \to \infty$, the fraction of usable habitats $h_t$ become independent of
the engineer density $e_t > 0$ (see eqs.\ (\ref{h}) and  (\ref{v}))  and so the equation for the engineer density (see eq.\  (\ref{e})) reduces
to the classic Ricker equation for which the carrying capacity is density independent. 

Overall we conclude that  increase of $\alpha$ and $\rho$ as well as decrease of $\delta$  favor the stability of the finite-engineers fixed point, but the determinant parameter is the intrinsic growth rate $r$. For instance, the finite-engineers fixed point $e^* > 0$  is always unstable for $r > 2$
and always stable for $ r < 1$.

\begin{figure}[!ht]
\centering
\includegraphics[width=0.48\textwidth]{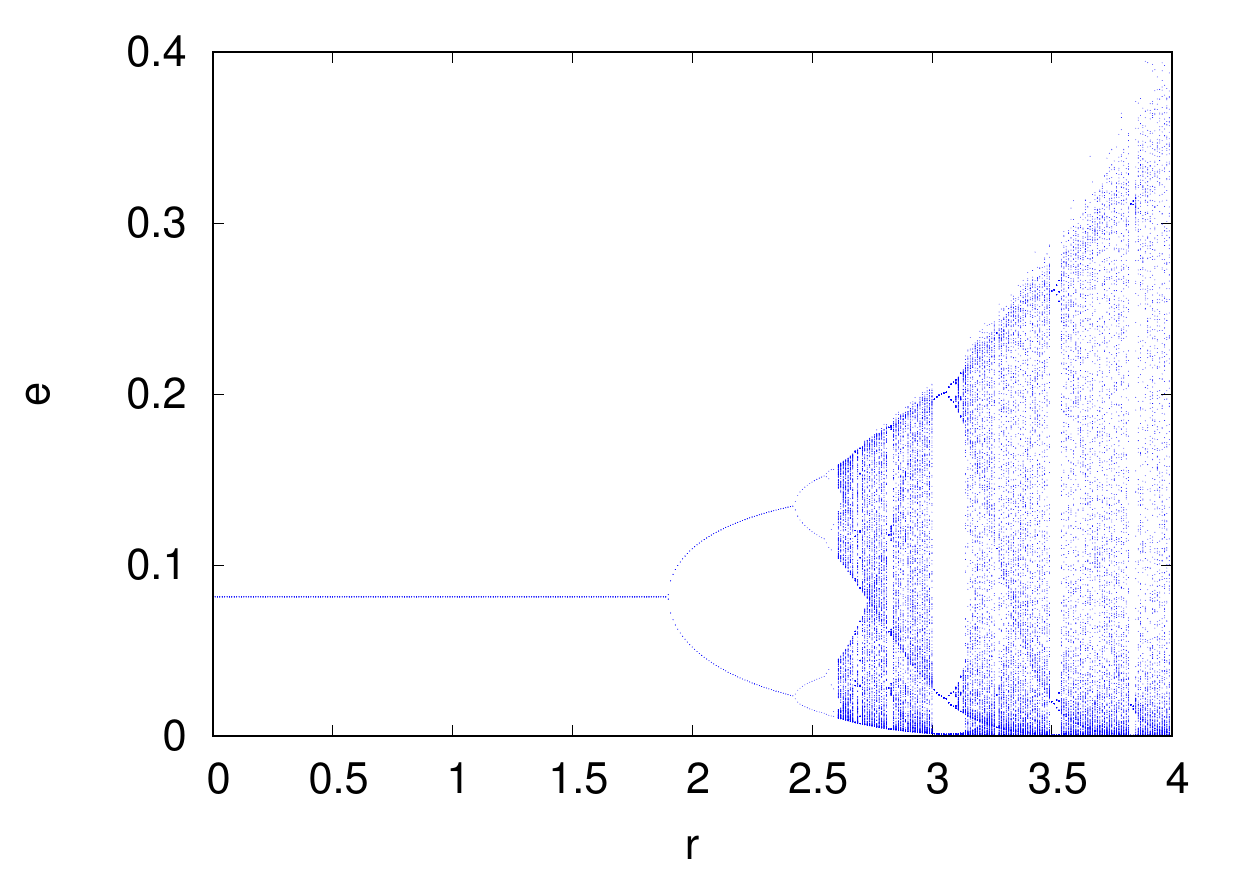}
\caption{(Color online) Bifurcation diagram for the local population dynamics  (\ref{e})-(\ref{v}) with  parameters  
$\alpha = 1$,  $\delta= 0.1$ and $\rho = 0.01$. The points on the y-axis show 
the values of the engineer density  visited asymptotically from all initial conditions with $e_0 > 0$.
 }
\label{fig:2}
\end{figure}
 
 \subsection{Bifurcation diagrams}\label{BD}
 
The analysis of the oscillatory regime for $r > r_c$ must be done by iterating numerically the equations of the  local population dynamics (\ref{e})-(\ref{v}). The  oscillatory regime comprehends  periodic oscillations and  chaos. A traditional way to present the results is through the so-called bifurcation diagrams \cite{Sprott_03} which exhibit a long-exposure photography of the attractors, as shown  in Fig.\ \ref{fig:2}. The bifurcation diagrams for the habitat fractions $v$ and $h$ show similar patterns.
  Since the source of nonlinearity of the population dynamics  is Ricker's formula for the growth of the engineer population  \cite{Ricker_54}, the observed period-doubling bifurcation cascade is expected. We note that the transitions  from fixed-points to two-point cycles  occur at $r = r_c$  (see Fig.\ \ref{fig:1}).

\begin{figure}[!ht]
\centering
\includegraphics[width=0.48\textwidth]{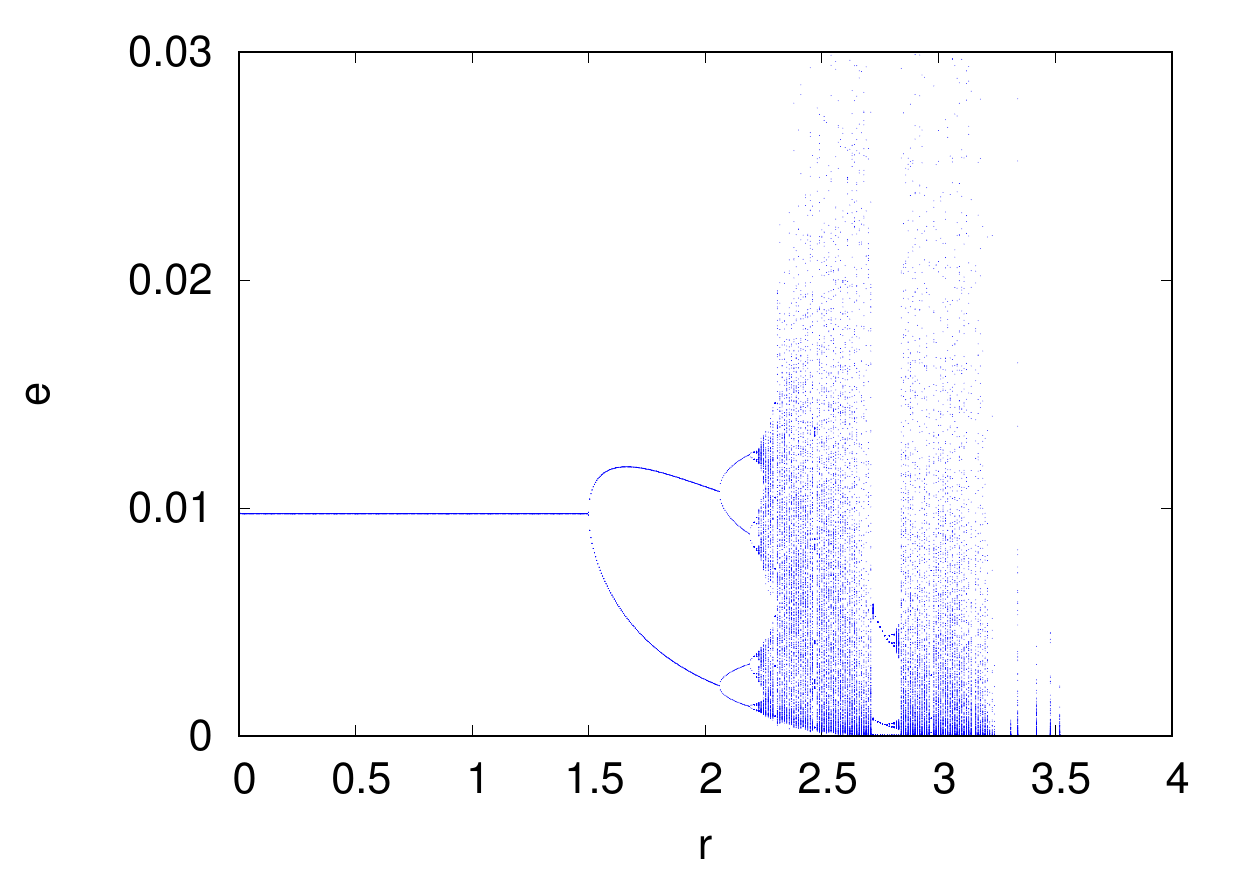}
\caption{(Color online) Same as for Fig.\ \ref{fig:2} but for the parameters 
$\alpha = 1$,  $\delta= 0.5$ and $\rho = 0.01$. For $r > 3.4$ the attractor has $e =h= 0$ with $x=e/h$  oscillating.
 }
\label{fig:3}
\end{figure}

There are, however, some interesting peculiarities about the dynamics  (\ref{e})-(\ref{v}) in the case the decay fraction $\delta$ is large, as shown in the bifurcation diagram of   Fig.\ \ref{fig:3}. The interesting point here  is that for large $r$ (more pointedly, $r>3.4$ for the  parameters of the figure) as well as for several small windows of the  interval of variation of $r$,  the attractor has $e=h=0$ and
$x = e/h$ oscillating, i.e., it is the same attractor that we came across in our study of the stability of the zero-engineers fixed point. We recall
that since $ \delta < \alpha$, the fixed point with $e^*=h^*=0$ is unstable.  
It is interesting that the domain of the zero-engineers oscillating attractor is practically unaffected  by the increase of the recovery fraction  $\rho$. In fact, setting $\rho = 1$ does not alter qualitatively the bifurcation diagram of  Fig.\ \ref{fig:3}. The presence of this  attractor for
large $r$ cannot  be explained by the violent oscillations and low minimum population densities that  may eventually lead to the extinction of the population and so justify the rarity of chaotic behavior in nature \cite{Thomas_80,Berryman_89}. In fact, for large $r$ the density $e$ does not experience large fluctuations and  simply oscillates with decreasing amplitude towards the zero-engineers attractor.
 We note that for $ \delta > \alpha$ the only attractors are the fixed point $e^*=h^*=0$  for $r < r_c$ and the attractor with $e=h=0$ and $x = e/h$ oscillating for  $r > r_c$, where $r_c$ is shown in Fig.\ \ref{fig:rc_e0}.

\section{Spatial population dynamics}\label{sec:spatial}

As described in Section \ref{sec:model} we allow that a   fraction $\mu$ of the engineer population in a given patch, say patch $i$,  moves to the  $K_i$ nearest neighbors of patch $i$.
The  density of engineers after the dispersal stage in each patch $i=1,\ldots, N$ is given by eq.\ (\ref{el}).
 In this section we  consider a chain with reflective boundary conditions 
 (i.e., $K_1=K_N = 1$ and $K_i = 2, \forall i \neq 1,N$)  and a square lattice with the Moore neighborhood  and reflective boundary conditions
  (i.e., $K_i=8$ for internal patches, $K_i = 5$ for patches on the edges and  $K_i = 3$ for patches on the corners of the lattice).  We recall that for internal 
  patches (or cells)  on a two-dimensional square lattice, the Moore neighborhood 
  is composed of a central patch and the eight patches that surround it.
 After dispersal of the engineers, the local population dynamics takes place within each patch:
\begin{eqnarray}
e_{i,t+1} &  =  & e'_{i,t} \exp \left [ r \left ( 1 - e'_{i,t}/h_{i,t} \right ) \right ]  \label{ei} \\
h_{i,t+1} & = & \left ( 1 - \delta \right ) h_{i,t} + c \left ( e'_{i,t} \right ) v_{i,t} \label{hi} \\
v_{i,t+1} & = &  \rho \left ( 1 - v_{i,t} -h_{i,t}  \right ) + \left [ 1- c \left ( e'_{i,t} \right ) \right ] v_{i,t}, \label{vi}
\end{eqnarray}
for $i=1, \ldots, N$. These equations together with eq.\ (\ref{el})  can be seen  as a coupled map lattice (see, e.g., \cite{Kaneko_92}) that describe the dynamics of the system of patches or metapopulation.

In the following, we  consider  a colonization or invasion scenario where at generation $t=0$ only the central patch $i_c$  of the lattice is populated, whereas the other patches are composed entirely of virgin habitats \cite{Comins_92}.

\subsection{One-dimensional lattice}

Here we consider a  chain with $N \geq 3 $ patches in order to study the colonization scenario.  Accordingly, we set  the initial density of the central patch  $e_{i_c,0}$ to  a random value drawn from a uniform distribution in the unit interval. In addition, we set $h_{i_c,0} = e_{i_c,0}$ and $v_{i_c,0} = 1 - h_{i_c,0} $. All the other patches have
$e_{i,0}=h_{i,0} = 0$ and  $v_{i,0} = 1$ for  $i \neq i_c$.  (Note that  $i_c = 2$ for $N=3$). 

\begin{figure}[!ht]
\centering
\includegraphics[width=0.48\textwidth]{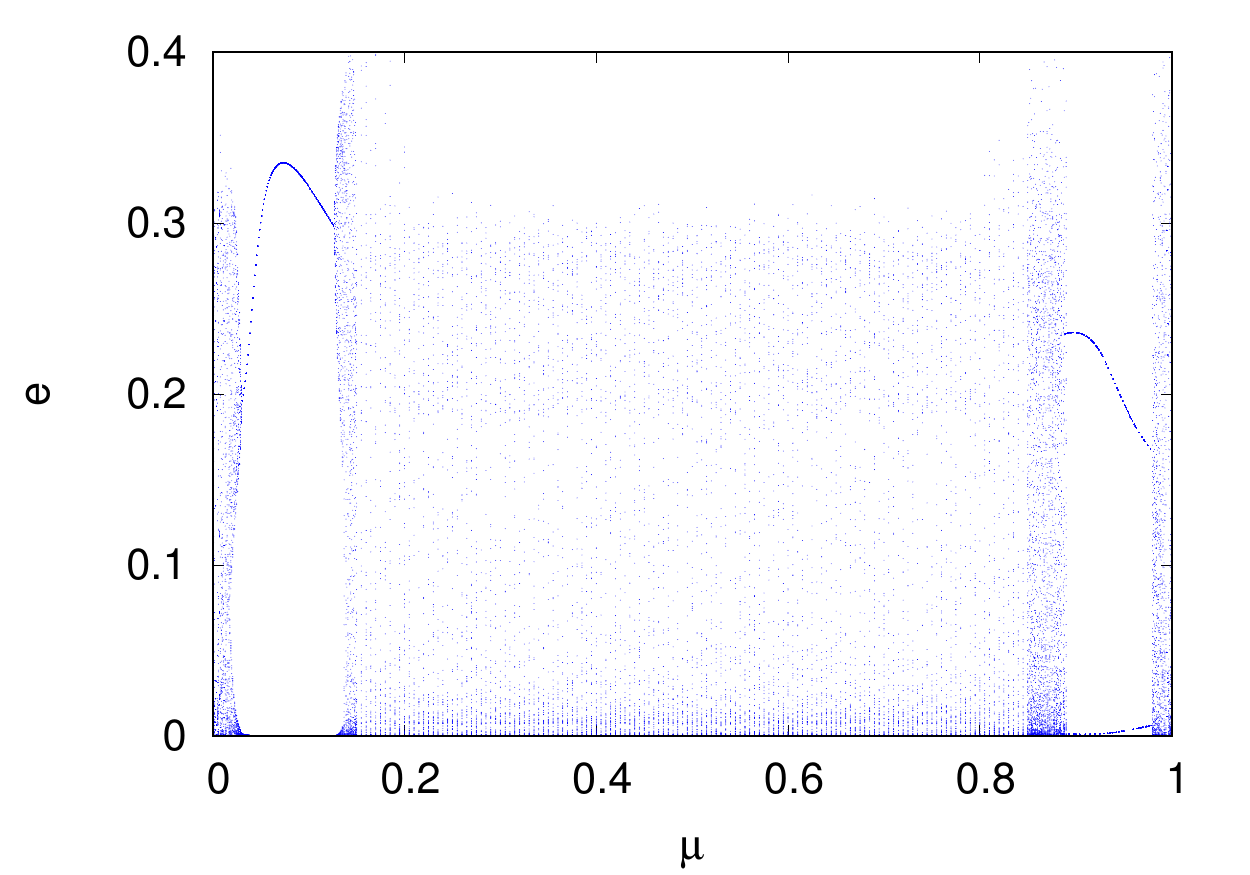}
\caption{(Color online) Bifurcation diagram for the engineer population in the central  patch  in  the case of $N=3$ patches  with  parameters  $r=3.6$, $\alpha = 1$,  $\delta= 0.1$ and $\rho = 0.01$. 
 }
\label{fig:N3}
\end{figure}

We find that the diffusive dispersal has no impact  on the behavior of the individual patches in the case the attractor of the local or single-patch dynamics is a fixed-point or an $n$-point cycle. In particular,  we find that the metapopulation is spatially homogeneous and oscillates with the same period of a single  patch. However, when the local dynamics is chaotic the  introduction of diffusive dispersal produces
remarkable results as  discussed next.  

\begin{figure}[!ht]
\centering
\includegraphics[width=0.48\textwidth]{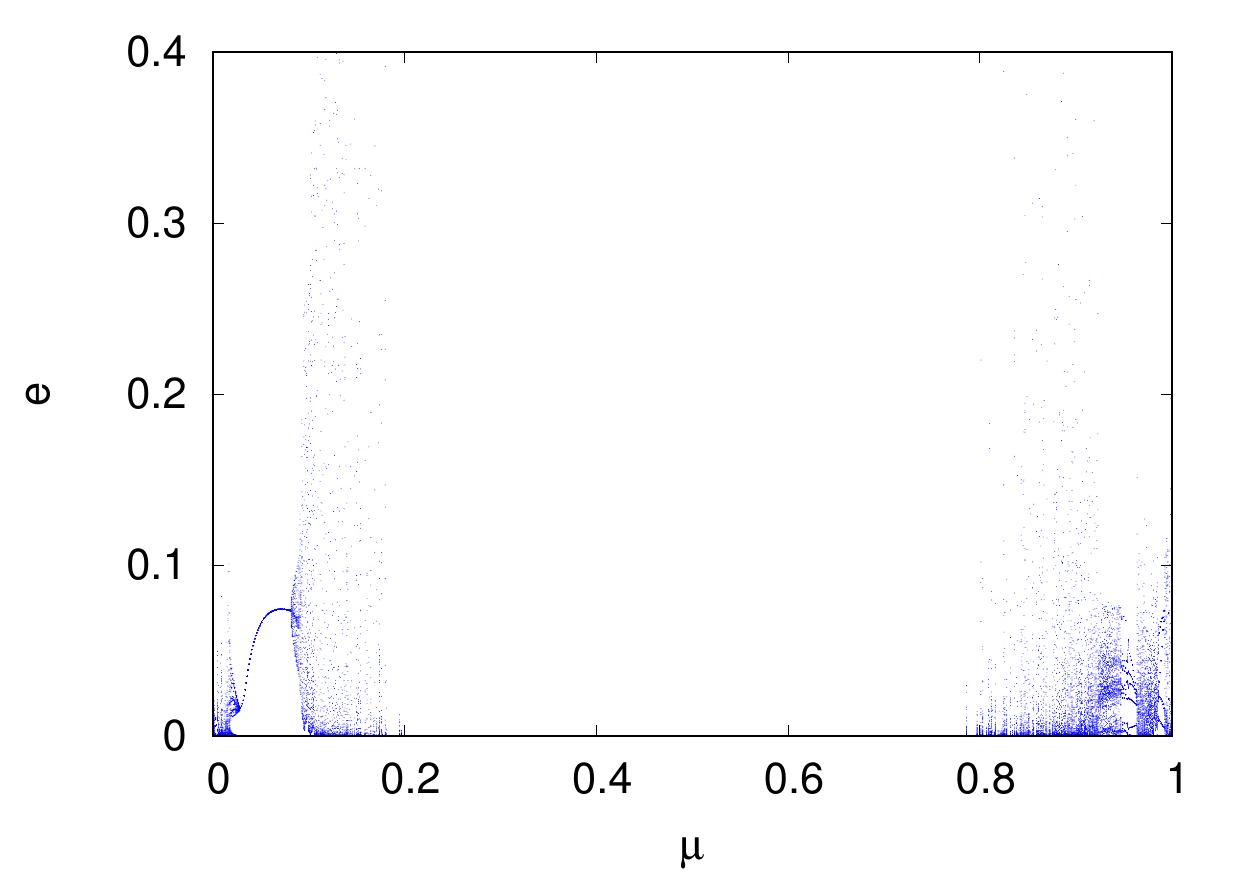}
\caption{(Color online) Bifurcation diagram for the engineer population in the central  patch  in  the case of $N=3$ patches  with  parameters  $r=3.6$, $\alpha = 1$,  $\delta= 0.5$ and $\rho = 0.01$. In the region  $0.2 < \mu  < 0.8$, as well as for $\mu=0$,  the sole  attractor is $e = h = 0$.
 }
\label{fig:N3_d05}
\end{figure}

The main effect of the diffusive dispersal is already visible in the case of $N=3$ patches as illustrated in Fig.\ \ref{fig:N3}, which shows 
the bifurcation diagram for the engineer population in  the central  patch of the chain  as function of the dispersal fraction $\mu$. The bifurcation 
diagrams are qualitatively similar  for  the other two  patches.  
This figure reveals that the chaotic behavior is suppressed  and the 
metapopulation enters a two-point cycle regime or a fixed-point regime for some values of $\mu$.
 For instance, in the region around $\mu = 0.1$ the three patches alternate between high ($e \approx 0.3$) and very low 
($e \sim 10^{-5}$) population densities, such that neighboring patches (i.e., patches 1 and 2 or patches 2 and 3) always have different densities. In the region around $\mu = 0.9$ the dynamics is attracted to a fixed point in which the two  unconnected patches have the same density.
We note that the reason the bifurcation diagram is not symmetric about $ \mu = 0.5$   is that the habitats do not move to neighboring   patches, only the engineers do  and when they arrive in a new patch they encounter a different environment.

Another unexpected effect of  the engineer dispersal is a partial avoidance of extinction. Consider the case illustrated in Fig.\ \ref{fig:3} where large  values of the intrinsic growth rate (more pointedly, $r > 3.4$) lead to the extinction of the engineers in the single-patch situation. The bifurcation diagram of Fig.\  \ref{fig:N3_d05} shows the effect of  dispersal for $r = 3.6$ and the same parameters used in Fig.\ \ref{fig:3}.  As in the single-patch case,  the only attractor is the zero-engineers attractor for a   large  range of values of  the dispersal fraction,
{\it viz}., for  $\mu \in \left [ 0.2, .8 \right ]$, but Fig.\  \ref{fig:N3_d05} shows that 
there are regions for small and large values of $\mu$ where the population of engineers can actually thrive due to the possibility of migration to more hospitable patches.

The bifurcation diagram for a large number of patches, say $N=101$, exhibits  a pattern similar to that for $N=3$ with a window of two-point cycles around $\mu =0.1$ However,  
 a large window of $n$-point cycles appears  around $\mu =0.5$ and the periodic window for large $\mu$  is suppressed. The spatial patterns observed within the periodic windows are essentially an juxtaposition of high and low population  densities. The next section will explore the more interesting spatial patterns that emerge in a two-dimensional lattice.
\captionsetup[subfigure]{labelformat=empty}
\begin{figure}[!ht]
\centering
\subfloat[t=100]{\includegraphics[width=0.48\textwidth]{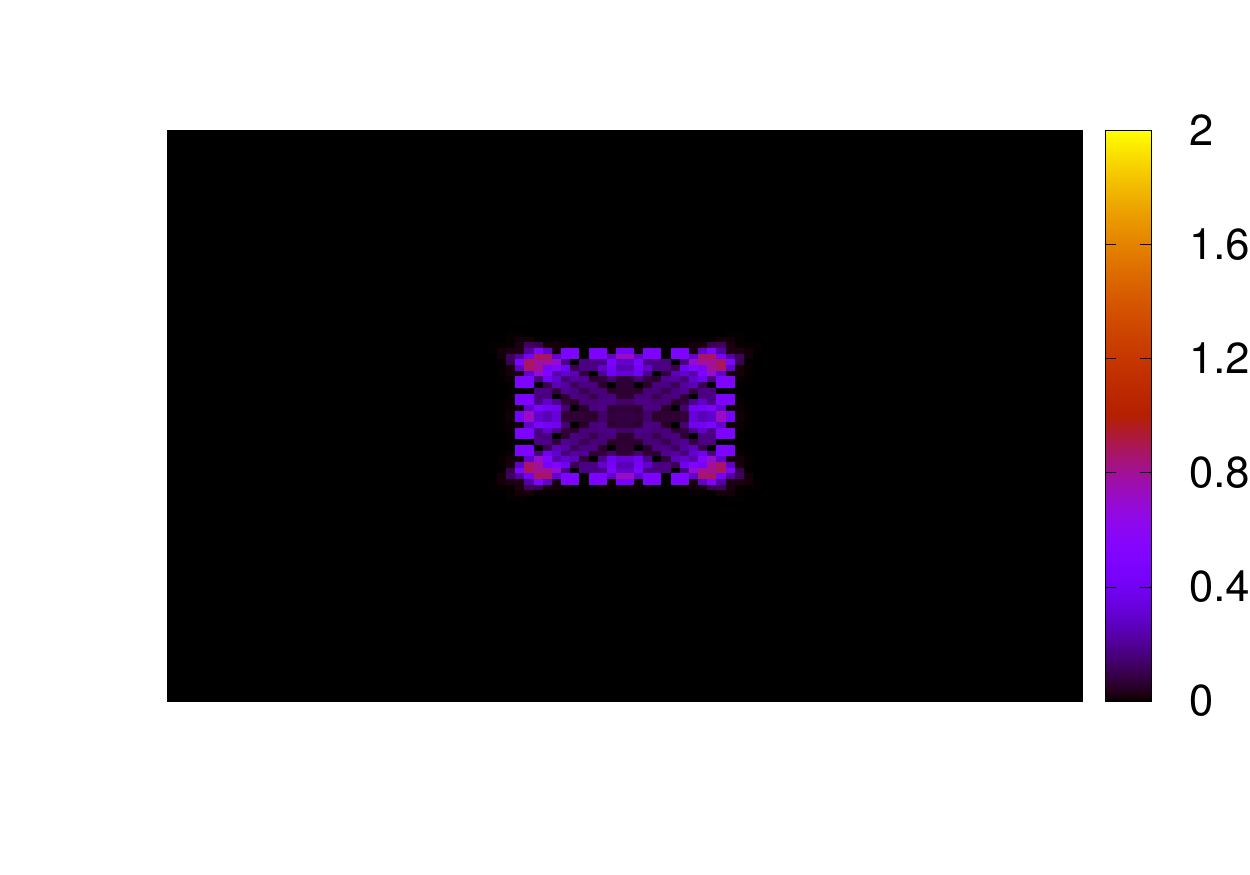}}\\
\subfloat[t=200]{\includegraphics[width=0.48\textwidth]{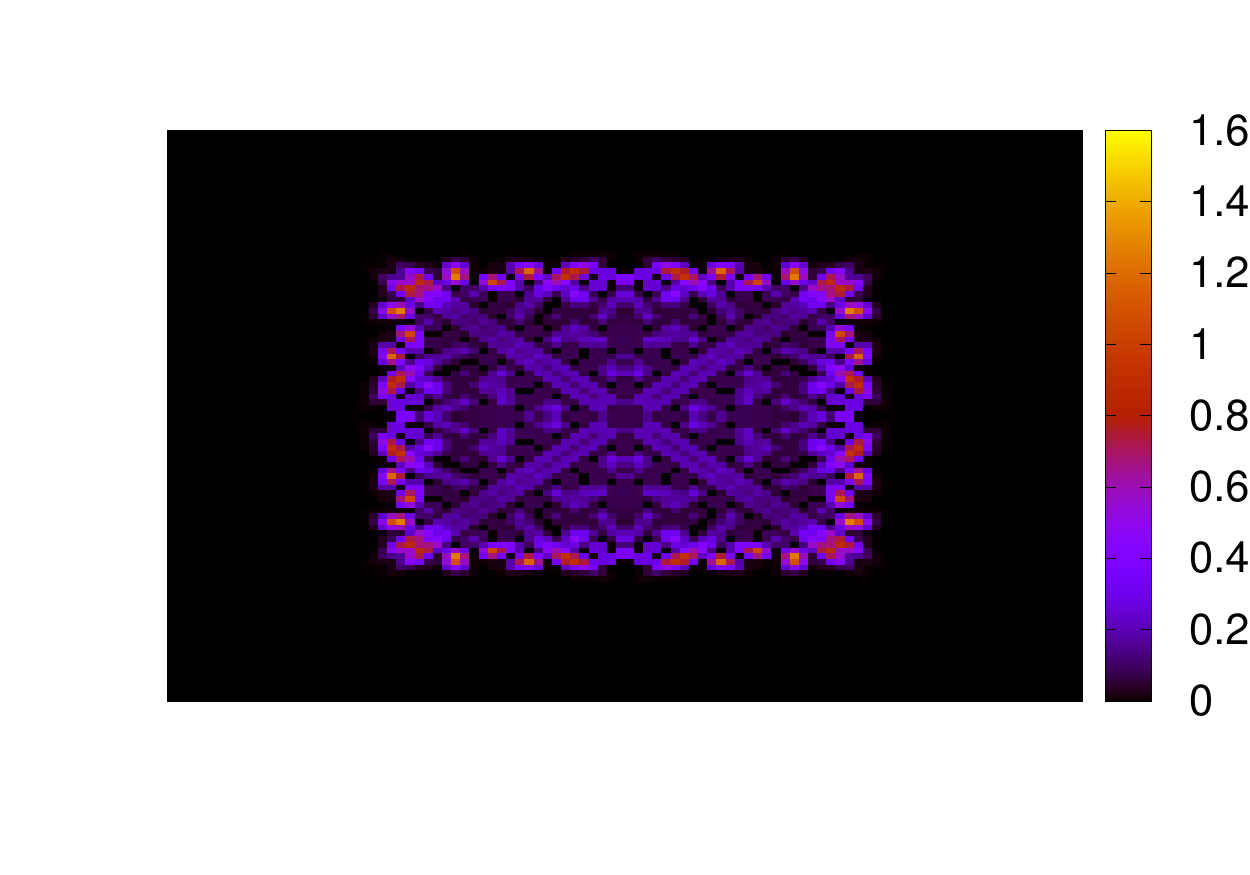}}
\caption{(Color online) Local density of engineers for  $101^2$ patches  arranged in a square lattice   at times $t=100$ and $t=200$ as indicated.
 In the initial condition, all patches are empty virgin habitats, except the central patch that has $e=h=v=0.5$. The model parameters are $\mu = 0.1$,  $r=3.6$, $\alpha = 0.5$, $\delta= 0.1$ and $\rho = 0.01$. The steady state is  the two-point cycle shown in the middle row of Fig.\ {\ref{fig:map01_2t}.}
 }
\label{fig:map01_t0}
\end{figure}

\captionsetup[subfigure]{labelformat=empty}
\begin{figure*}[!ht]
\centering
\subfloat[t] {\includegraphics[width=0.48\textwidth]{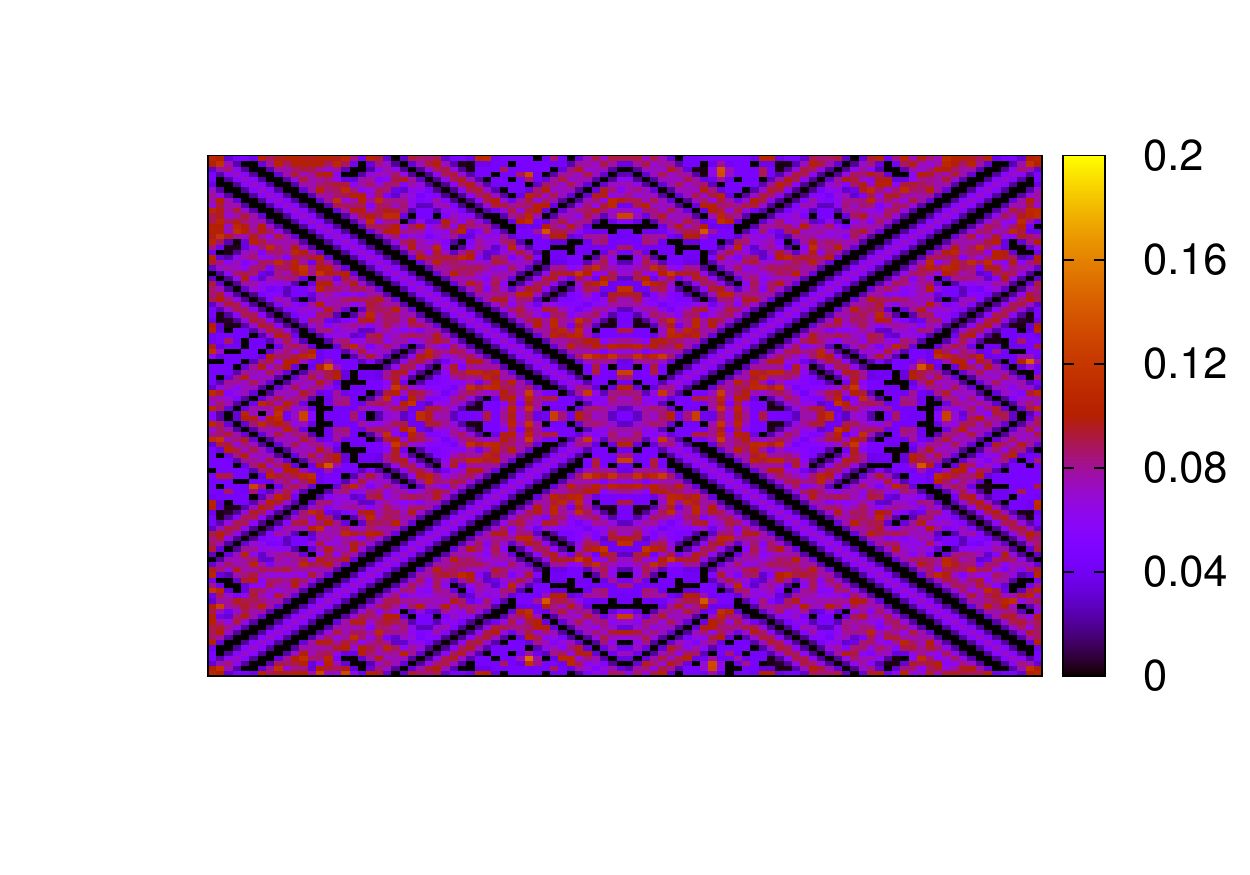}}
\subfloat[t+1]{\includegraphics[width=0.48\textwidth]{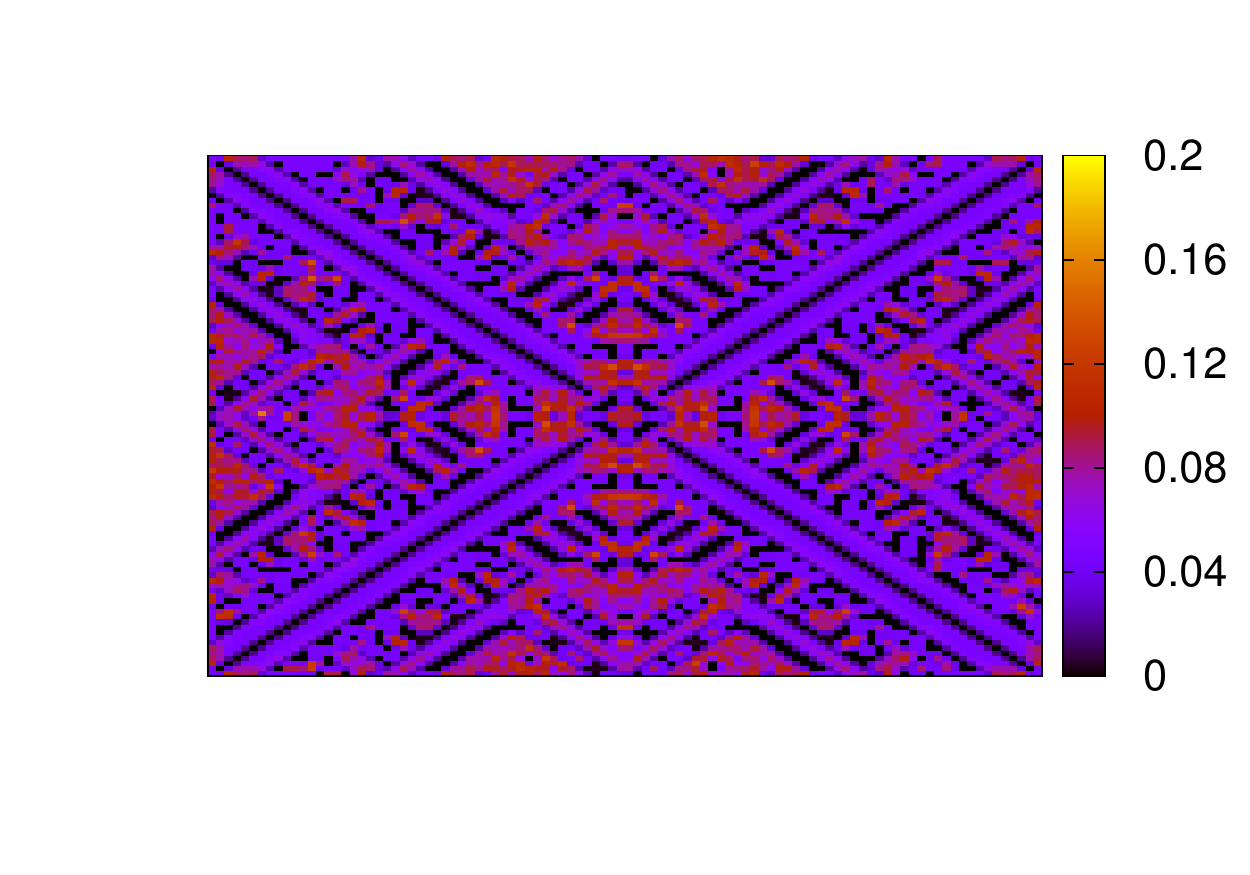}} \\
\subfloat[t] {\includegraphics[width=0.48\textwidth]{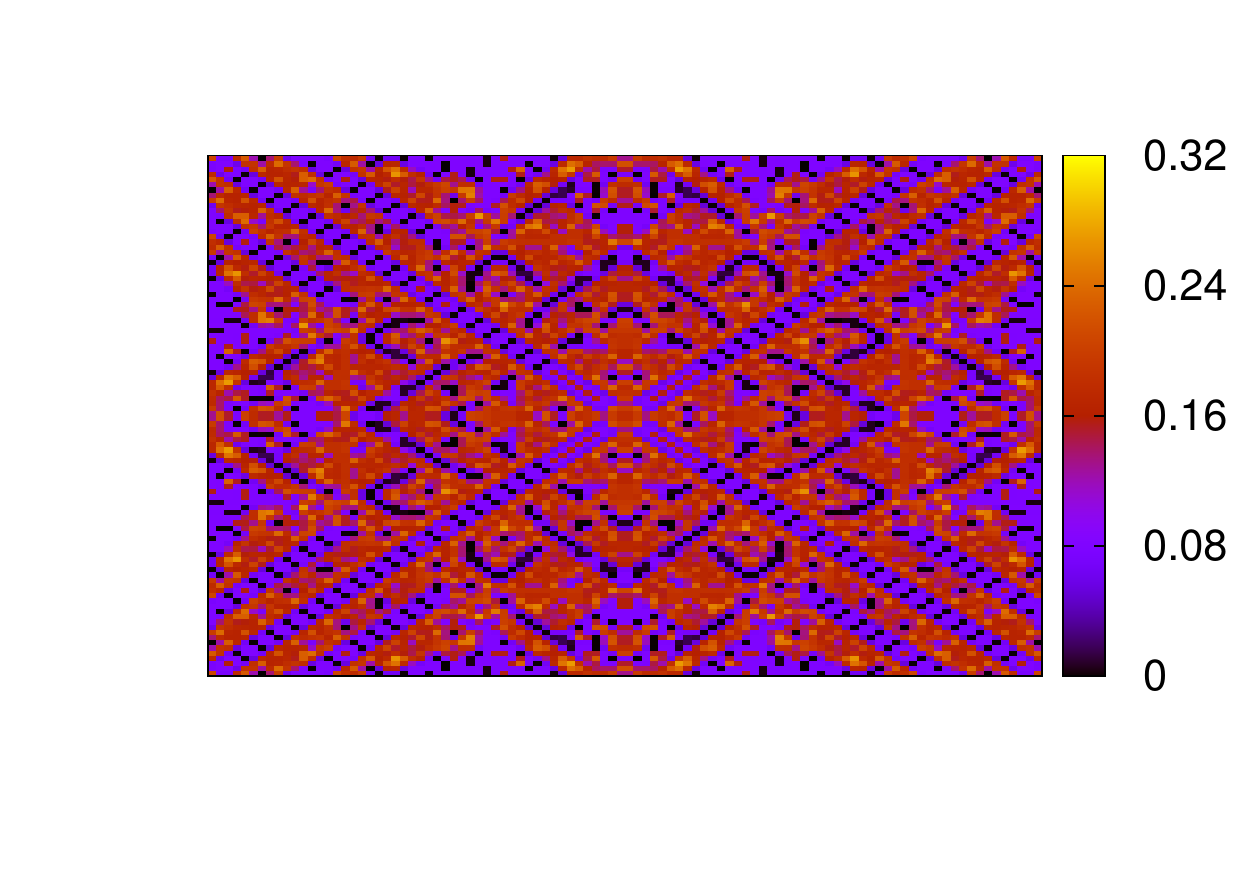}}
\subfloat[t+1] {\includegraphics[width=0.48\textwidth]{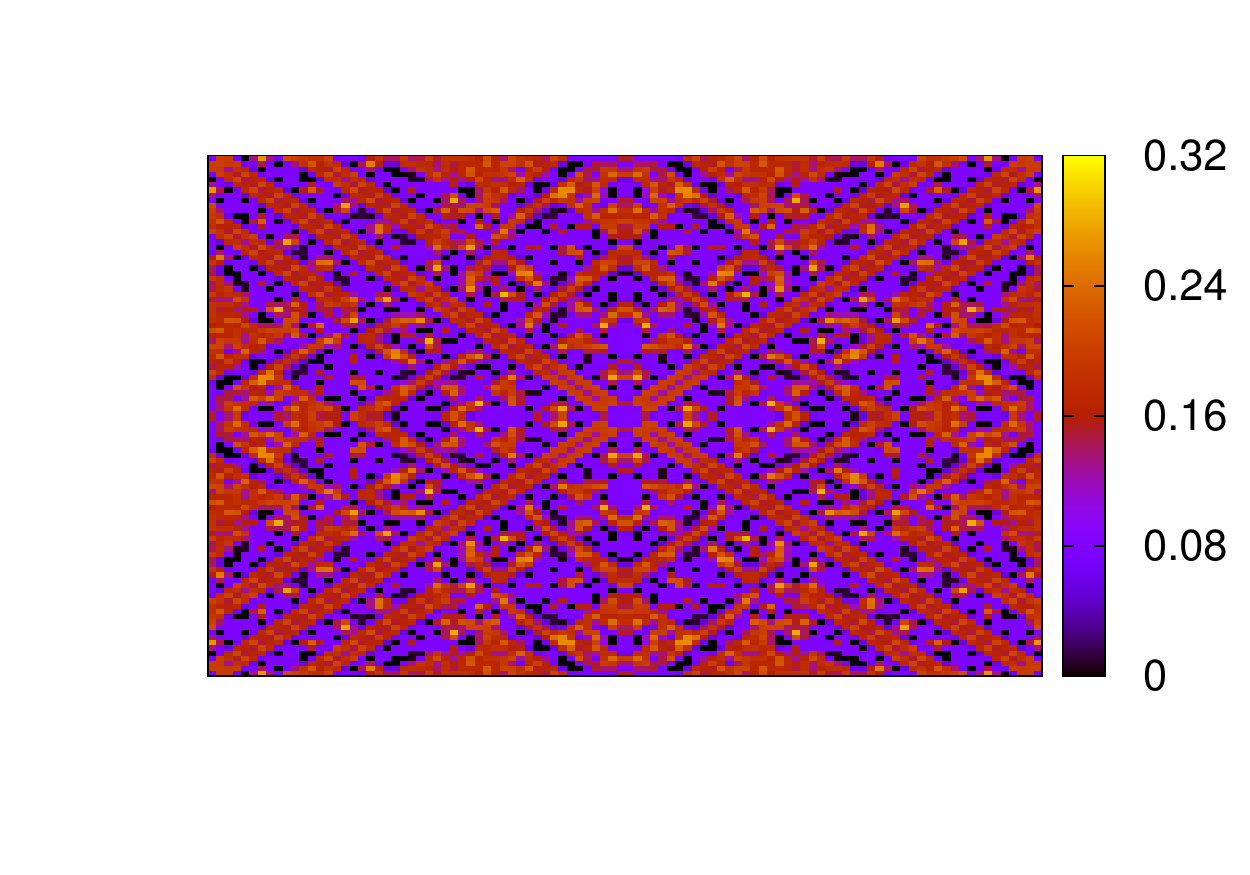}}\\
\subfloat[t] {\includegraphics[width=0.48\textwidth]{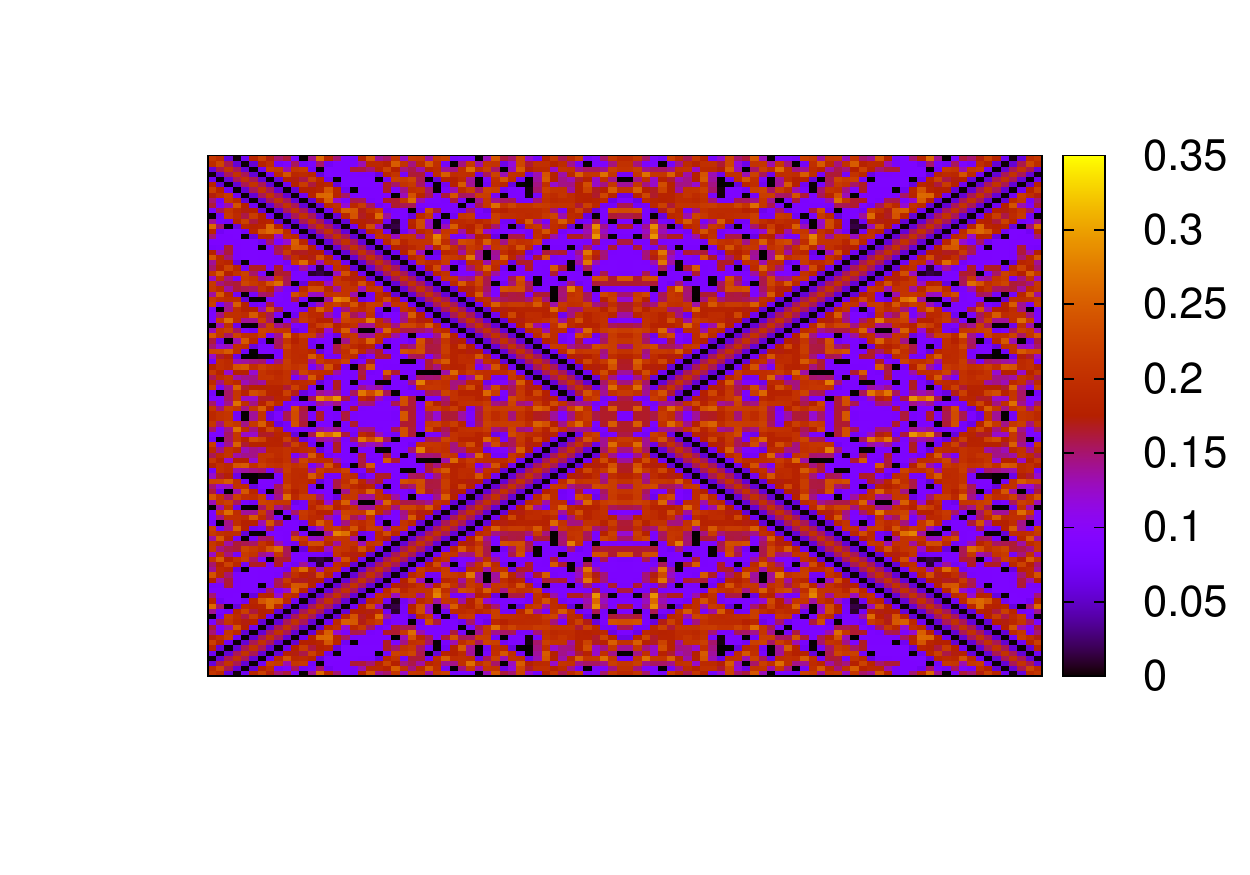}}
\subfloat[t+1] {\includegraphics[width=0.48\textwidth]{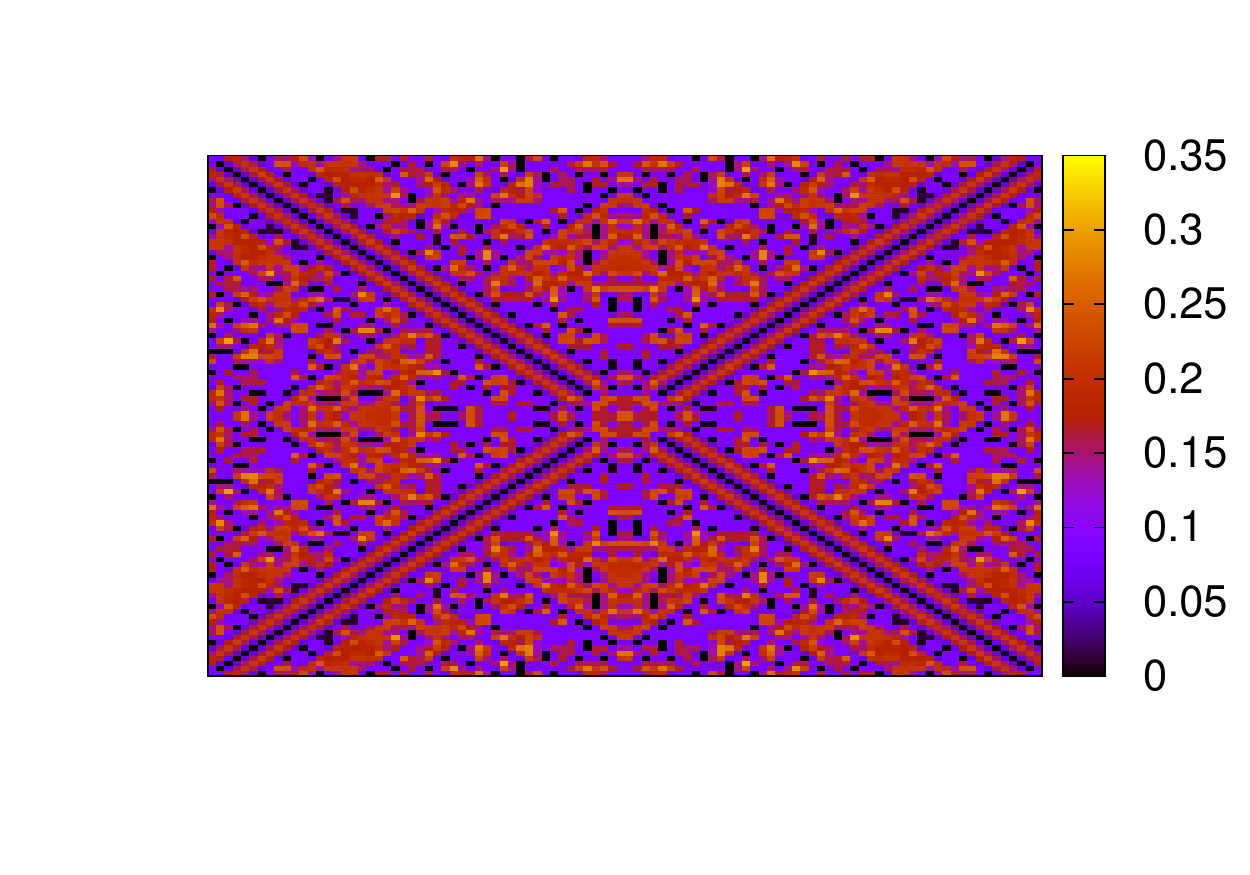}}
\caption{(Color online) Local density of engineers for  $101^2$ patches  arranged in a square lattice  in the  two-point cycle steady state for
$\alpha =0.11$  (upper row), $\alpha = 0.5$ (middle row)  and $\alpha = 1.0$ (lower row)  The panels in a same row  show the lattice at consecutive times. In the initial condition, all patches are empty virgin habitats, except the central patch that has $e=h=v=0.5$. The  other parameters are $\mu = 0.1$,  $r=3.6$, $\delta= 0.1$ and $\rho = 0.01$. 
 }
\label{fig:map01_2t}
\end{figure*}
   
 \subsection{Two-dimensional lattice}

 As expected, the   conclusion that diffusive dispersal  impacts the metapopulation only when the single-patch dynamics is chaotic applies to the square lattice as well. Its effect is the appearance of windows in the bifurcation diagrams   in which chaos reverts to periodic motion.

\captionsetup[subfigure]{labelformat=empty}
\begin{figure*}[!ht]
\centering
\subfloat[t=50]{\includegraphics[width=0.48\textwidth]{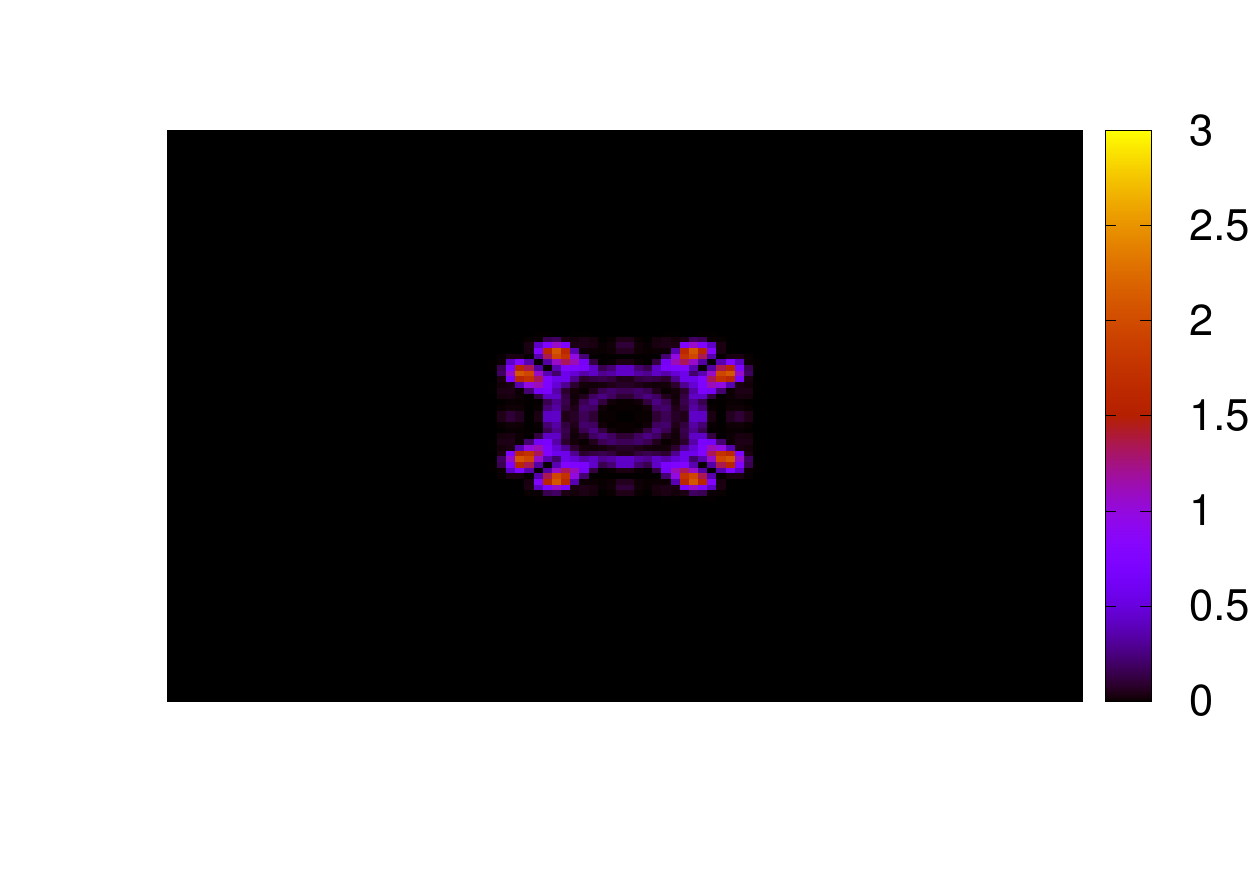}}
\subfloat[t=150]{\includegraphics[width=0.48\textwidth]{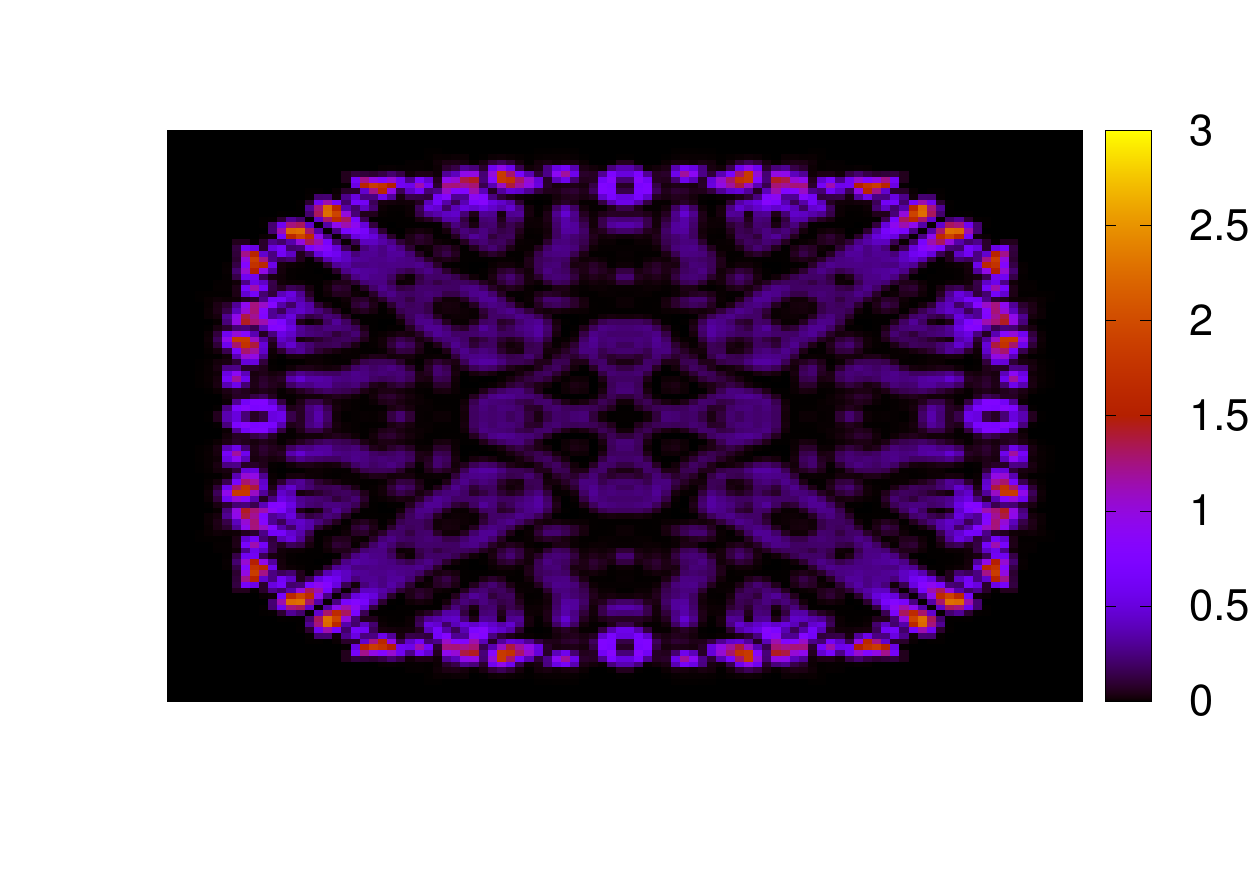}}\\
\subfloat[t=240]{\includegraphics[width=0.48\textwidth]{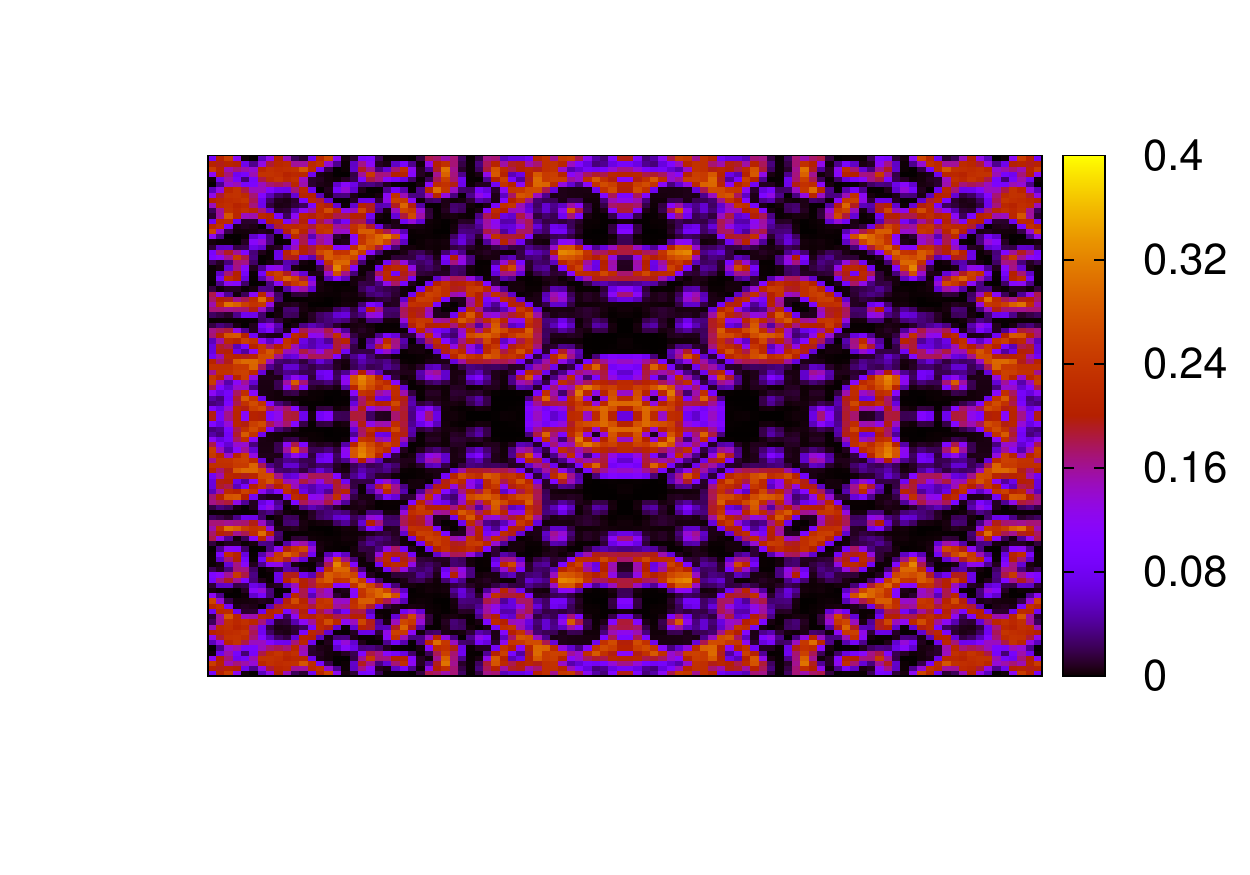}}
\subfloat[t=250]{\includegraphics[width=0.48\textwidth]{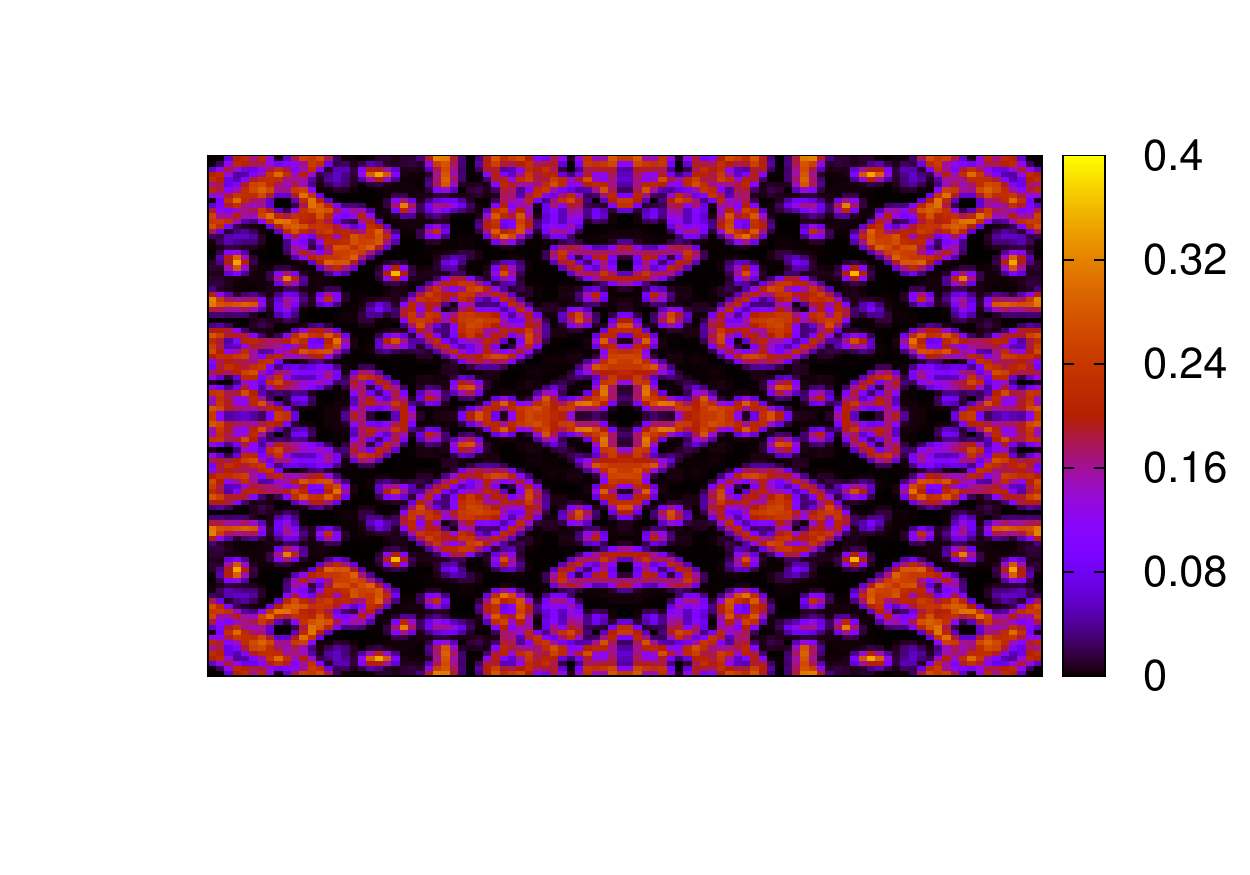}}\\
\subfloat[t=260]{\includegraphics[width=0.48\textwidth]{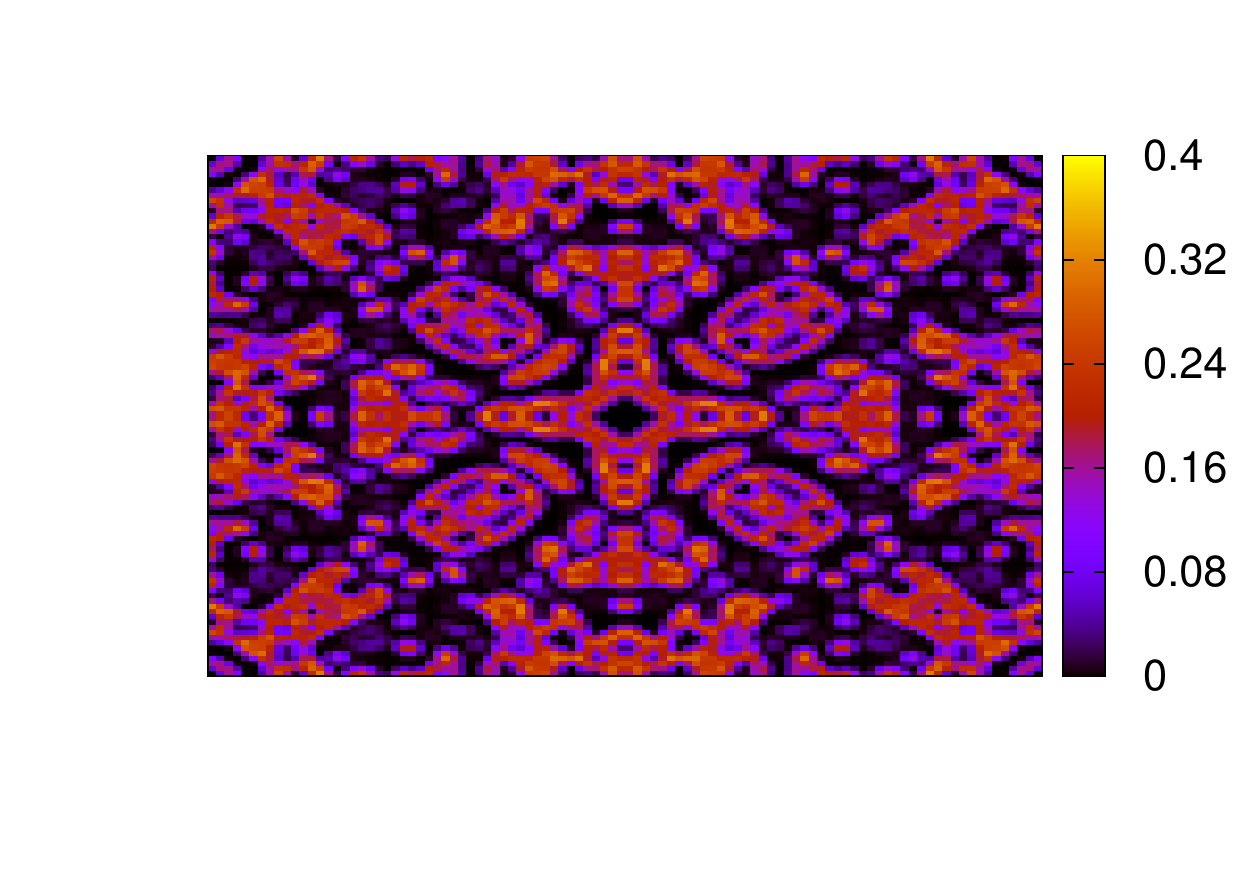}}
\subfloat[t=270]{\includegraphics[width=0.48\textwidth]{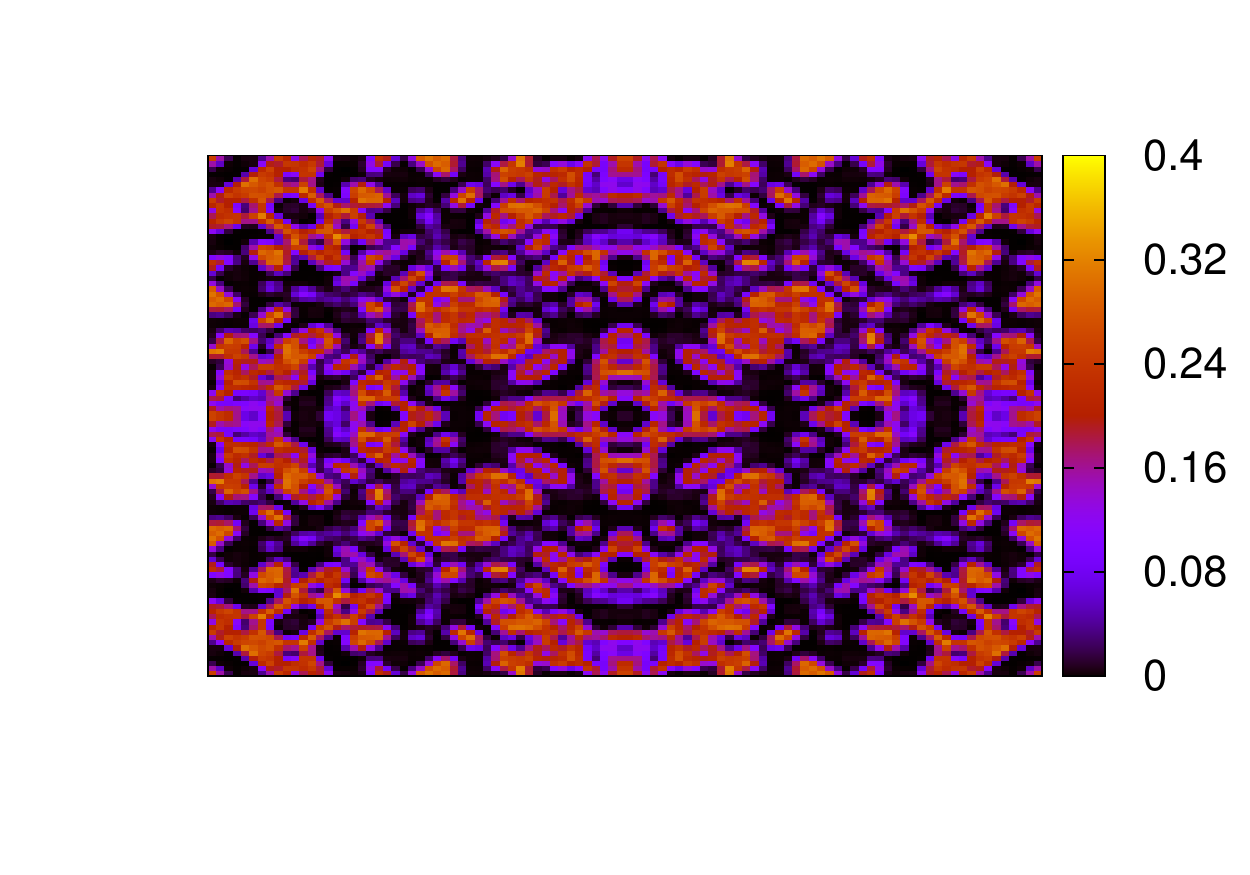}}
\caption{(Color online) Local density of engineers for  $101^2$ patches  arranged in a square lattice  in the spatio-temporal chaotic regime.
 The panels   show the lattice at  different  times as indicated. In the initial condition, all patches are empty virgin habitats, except the central patch that has $e=h=v=0.5$. The model parameters are $\mu = 0.5$,  $r=3.6$, $\alpha = 1$, $\delta= 0.1$ and $\rho = 0.01$. 
 }
\label{fig:map05}
\end{figure*}

 Since the numerical errors accumulate very fast when we implement the local dynamics of thousands of  interconnected patches in the chaotic regime, in spite of  the  use of quadruple precision, it is useful to have a  marker to signal when the numerical results are unreliable. Since we start the dynamics with a single
populated  patch located at the center of the lattice, the spreading of the engineers must preserve the symmetries of the square lattice. Hence, in the colonization scenario
our marker of the unreliability of the numerical results is the breaking of the symmetries of the square lattice. We refer the reader to Ref.\ \cite{Rodrigues_15} for a detailed discussion of
the symmetry breaking caused by the chaotic amplification of numerical noise.
   
Figure \ref{fig:map01_t0} shows two snapshots of the lattice at short times, when the spreading colony is still far from the borders. The figure reveals two interesting features, {\it viz}.,
the high density of engineers in the patches located at the wavefront  and the square symmetry of the colony due to the choice of the Moore neighborhood. For  low values of the dispersal fraction (e.g., $\mu = 0.1$) the dynamics converge to a  two-point cycle steady state, and 
 Fig.\ \ref{fig:map01_2t}  shows snapshots at two consecutive times in this state.  The rich tapestry of these patterns  is the result of  the propagation of arrow-shaped groups of migrants out of the lattice center that  form before the  wavefront reaches the lattice borders, as shown in  Fig.\  \ref{fig:map01_t0}. We find essentially the same spatial patterns for different values of $\delta$ and $\rho$  provided the dynamics enters the two-point cycle stationary regime.

A more interesting scenario  appears in the spatio-temporal chaotic regime of the  metapopulation  that  occurs for values of the dispersal fraction outside the periodic windows. Figure \ref{fig:map05} shows snapshots of the lattice at different times for  $\mu=0.5$. The new feature revealed by these patterns is the existence of  isolated  isles of engineers surrounded by patches of virgin habitats. This patchiness, where regions of
high and  low population densities alternate, is characteristic of  biological patterns \cite{Levin_76,Greig_79}.

\begin{figure}[!ht]
\centering
\includegraphics[width=0.48\textwidth]{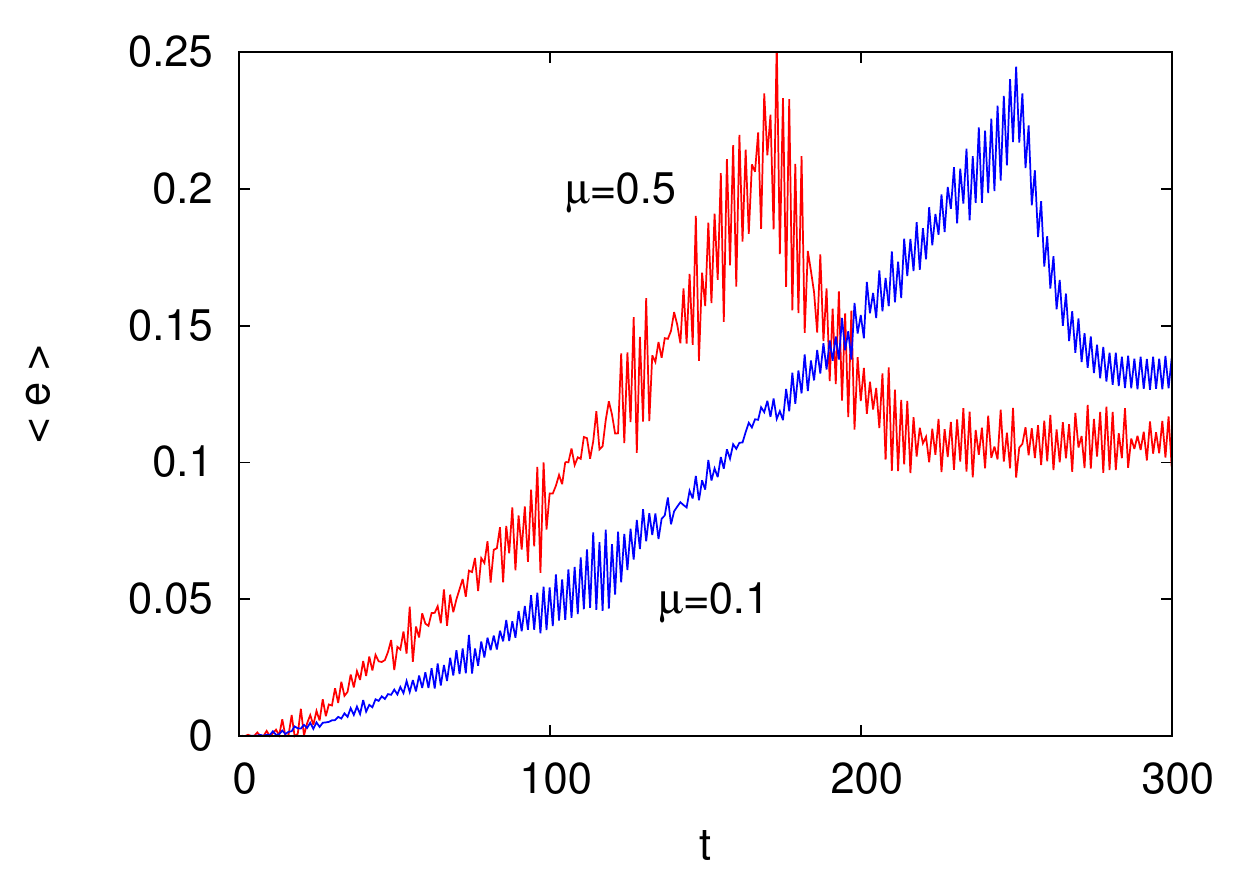}
\caption{(Color online) Time evolution of the mean engineer density $\langle e \rangle$ in a square lattice with $101^2$ patches  for the dispersal fractions $\mu=0.1$ and $\mu = 0.5$ as indicated. The other model parameters are   $r=3.6$,
$\alpha = 1$,  $\delta= 0.1$ and $\rho = 0.01$. 
 }
\label{figav}
\end{figure}

It is interesting that although  the patterns shown in Figs.\ \ref{fig:map01_2t}  and \ref{fig:map05} differ in their dispersal fraction only,  the mean density of engineers in the metapopulation 
$\langle e_t \rangle = \sum_{i=1}^N e_{i,t}/N$ at the stationary  regime decreases with increasing $\mu$ as shown in Fig.\ \ref{figav}. The initial  increase of $\langle e_t \rangle $  reflects the expansion phase of the engineers that halts when  they reach the lattice borders.  The end of  the availability of unexplored virgin habitats leads to a sharp drop on the density of engineers, which then enters the stationary regime  with oscillations of limited amplitude. This phenomenon was also observed  in the study of the one-dimensional lattice. In fact,  the patches at the borders of the colonization wavefront exhibit a very high density of engineers $e \approx 3.0$ (see panels for $t=50$ and $t=150$ in Fig.\ \ref{fig:map05})  that are sustained  by the unexplored patches  ahead of the wavefront. When the supply of unexplored virgin habits is exhausted, the values of the highest densities undergo an  almost tenfold  drop and   plunge to $e \approx 0.4$.  
We note that Fig.\  \ref{figav} offers a crude way to estimate the mean speed of the colonization wavefronts as  the ratio between the shortest distance from the center to the borders  of the lattice and the time to reach those borders.

Finally, we should mention that we do not find any significant variations in the spatial  patterns for different  lattice sizes. In fact, as pointed out before the patterns close to the center of the lattice are formed well before the engineers reach the lattice boundaries. 
Hence the particular 
 geometric patterns observed in the square lattice are essentially a consequence of  the choice of the  Moore neighborhood for the engineers dispersal as well as of the symmetry of the
 initial distribution of the engineers in the patches \cite{Mistro_12}.

 \section{Discussion}\label{sec:conc}
 
 There are hardly any ecosystems on earth  that have not been engineered by past or present organisms, as attested by the highly  unstable, life-regulated composition of  the atmosphere of our planet \cite{Margulis_74}. However, the importance of ecosystem engineering \cite{Jones_94} or, more generally, of niche construction \cite{Odling_03} has been somewhat overlooked by  the  ecological and evolutionary literatures. 
(We refer the reader to the ongoing debate \cite{Scott_13,Laland_14} within the evolutionary biology  community  whether  niche construction is a new concept on evolution or  is simply  the  well-known feedback between organisms and the environment, whose study  was pioneered  by Darwin himself \cite{Darwin_81}. See also \cite{Mayr_01}  for support of the view that theories in  evolutionary  biology  are based  on  
concepts  rather  than  laws.)
We note that within the  theoretical  ecology perspective,  the population dynamics of predator-prey systems has been  studied for almost a century \cite{Lotka_25,Volterra_26}, whereas a first attempt to  model mathematically the dynamics of ecosystem engineers took place  in the mid 1990s only \cite{Gurney_96}. The interesting feature of the ecosystem engineers model  proposed  by Gurney and  Lawton  is that both the density of organisms -- the engineers -- and the quality of their habitats vary in time, as their survival depends on the existence of engineered   habitats \cite{Gurney_96}. This contrasts with  less transparent models of ecosystem engineers where the habitat changes are not explicitly taken into account  \cite{Cuddington_09}.

Although there are a few unquestionable examples of ecosystem engineers, such as  beavers that carry out extensive changes on their ecosystems through clearcutting and damming 
\cite{Wright_02} and caterpillars that create shelters from leaves that may also be occupied by other organisms \cite{Jones_97}, the field studies of these model systems are not complete
enough to parameterize the theoretical models \cite{Gurney_96}. We note that, in addition to the usual demographic variables, those models require information about the transition rates between the different habitat types. Perhaps, the study of the ultimate ecosystem engineers, humans \cite{Smith_07},  may offer the desired data: primitive societies that rely on  slash-and-burn and shifting cultivation agricultural systems  as their means of subsistence \cite{Waters_07}  may be seen as  model systems of spatial ecosystem engineers, since they must leave  their burnt fields and return back after they recover in a cyclic scheme similar to that introduced in our spatial model.

A word is in order about the existence of a natural spatial scale in our model of ecosystem  engineers. We envision two scenarios to which the spatial model could be applied: the colonization of an archipelago and the colonization of a galaxy.  Regarding the latter scenario, we remind the reader of the concept of terraforming as a planetary engineering process. In both cases, the engineers are humans, which can be viewed as the ultimate ecosystem engineers as  pointed out before, but the spatial scales are completely distinct. The common feature is that the typical size of the patches (isles or planets) is much smaller than the typical distance between patches. In fact, an implicit assumption of the model of Gurney and  Lawton \cite{Gurney_96} is that the engineers move between habitats within a same patch  with virtually infinite speed,  otherwise the virgin habitats would remain virgin forever.
Hence the relevant condition for  the applicability  of the model is that the diffusion rate within a patch be much greater than the diffusion rate among patches. This condition may  arise from differences in the spatial scales, as already mentioned, or from the lack of an efficient  technology to  travel between patches.

As expected, we find that  the local or single-patch dynamics converges to zero-engineers attractors whenever the number of  usable habitats decreases more rapidly than they are produced by a vanishingly small population of engineers working on virgin habitats, i.e., whenever the decay fraction $\delta$ is greater than the per capita productivity $\alpha$. A curious feature of the local dynamics, which is  due to the density-dependent carrying capacity in Ricker  equation,  is the existence of an oscillating zero-engineers attractor for which both the density of engineers $e_t$ and the fraction of usable habitats $h_t$ tend to zero in the time-asymptotic limit, but the ratio $x_t = e_t/h_t$ oscillates with finite amplitude.
Interestingly, the appearance of this attractor for large values of the engineers' intrinsic growth rate $r$  leads to the extinction of the population even in an apparently thriving scenario where  $\alpha > \delta$. Otherwise, the local dynamics exhibits the period-doubling route to chaos, which is expected since the population growth  is  governed by Ricker's model \cite{Ricker_54}.

The engineers' intrinsic growth rate $r$ is the parameter responsible for the main differences  between the time-discrete and the time-continuous single-patch dynamics.  In fact,  in the continuous version, $r$ simply sets the natural time-scale and so it plays no role at all in the dynamics \cite{Gurney_96}. Similarly,  the value of $r$ is inconsequential for  the calculation of the finite-engineers fixed point of the time-discrete model, but it is crucial for establishing its local  stability  (see Section \ref{subsec:FE}). In addition, only large values of $r$ can sustain  the chaotic dynamical behavior, which is  the most distinctive  feature of the time-discrete formulation of the population dynamics of ecosystem engineers.

We find that the spatial organization of  patches  and the dispersal of a fraction $\mu$ of engineers to neighboring patches have no influence on the metapopulation dynamics, provided that the attractors of the local dynamics are periodic.  In fact, our simulations show that the finite-engineers fixed point and the $n$-point cycles are always stable with respect to spatially inhomogeneous perturbations. However,  in the  case  the local dynamics is chaotic  the diffusive dispersal produces nontrivial effects, as shown in Figs.\  \ref{fig:map01_t0}, \ref{fig:map01_2t} and \ref{fig:map05}.  In particular, for certain values of the dispersal fraction $\mu$ the chaotic behavior  is suppressed and the dynamics enters  a two-point cycle or a fixed-point attractor (see Fig.\ \ref{fig:N3}). Perhaps more telling is
the finding that diffusive dispersal can prevent  the extinction of the metapopulation in a case where a single population  would die off in the absence of dispersal (see  Fig.\ \ref{fig:N3_d05}). This finding is in agreement with the notion that  extinctions resulting from  large fluctuations of the chaotic dynamics  can be eluded if the population  is composed of patches weakly coupled by migration   \cite{Allen_93}. 

Our more interesting findings  regarding the spatial aspects of the model of ecosystem  engineers are the patchy,  biological-like  patterns  illustrated in Fig.\ \ref{fig:map05}, which appear only in the regime of spatio-temporal chaos of the metapopulation dynamics. Since  the chaotic behavior is often viewed as a mathematical artifact rather than a genuine property of  ecological systems \cite{Berryman_89,Hassell_76}, it may be argued that our results  suggest the absence of patchy patterns in  natural engineer ecosystems. However, since the chaotic dynamics emerge when positive feedback growth processes are reinforced and  regulatory (negative feedback) processes are delayed,  and since these processes may be purposely altered  by the engineers actions \cite{Berryman_89}, the emergence of
the chaotic dynamics, and  consequently  of the nontrivial spatial organization, may be  commonplace in engineered ecosystems.

\acknowledgments
The research of JFF was  supported in part by grant
15/21689-2, S\~ao Paulo Research Foundation
(FAPESP) and by grant 303979/2013-5, Conselho Nacional de Desenvolvimento 
Cient\'{\i}\-fi\-co e Tecnol\'ogico (CNPq).  CF  was supported by grant 15/21452-2, S\~ao Paulo Research Foundation (FAPESP).



\end{document}